\documentclass[12pt]{iopart}

\usepackage{hyperref, multirow, color, bbold, textcomp}
\usepackage[dvipsnames]{xcolor}
\usepackage{orcidlink}

\usepackage{tikz, varwidth}
\usetikzlibrary{calc,positioning,shapes,shapes.geometric,shapes.misc,
  shadows,arrows,arrows.meta}

\newcommand{\response}[1]{{\color{black}{#1}}}

\usepackage[normalem]{ulem} %for \sout

\newcommand{\Fullerton}{Nicholas and Lee Begovich Center for Gravitational-Wave
  Physics and Astronomy, California State University Fullerton, Fullerton, CA
  92834, USA} \newcommand{\FullertonId}{1}

\newcommand{\Caltech}{Theoretical Astrophysics 350-17, California Institute of
  Technology, Pasadena, CA 91125, USA} \newcommand{\CaltechId}{2}

\newcommand{\CornellPhysics}{Department of Physics, Cornell University, Ithaca,
  New York 14853, USA} \newcommand{\CornellPhysicsId}{3}

\newcommand{\CornellLepp}{Laboratory for Elementary Particle Physics, Cornell
  University, Ithaca, New York 14853, USA} \newcommand{\CornellLeppId}{4}

\newcommand{\CornellCcaps}{Cornell Center for Astrophysics and Planetary
  Science, Cornell University, Ithaca, New York 14853, USA}
\newcommand{\CornellCcapsId}{5}

\newcommand{\Northwestern}{Center for Interdisciplinary Exploration and
Research in Astrophysics, Northwestern University, Evanston, IL 60201, USA}
\newcommand{\NorthwesternId}{6}

\newcommand{\Icts}{International Centre for Theoretical Sciences, Tata Institute
  of Fundamental Research, Bangalore 560089, India} \newcommand{\IctsId}{7}

\newcommand{\Aei}{Max Planck Institute for Gravitational Physics (Albert
  Einstein Institute), D-14467 Potsdam, Germany} \newcommand{\AeiId}{{8}}

\newcommand{\UCLA}{Mani L.~Bhaumik Institute for Theoretical Physics, Department
of Physics and Astronomy, UCLA, Los Angeles, CA 90095} \newcommand{\UCLAId}{9}

\newcommand{\Oberlin}{Department of Physics and Astronomy, Oberlin College,
  Oberlin, Ohio 44074, USA} \newcommand{\OberlinId}{{10}}

\newcommand{\WSU}{Department of Physics \& Astronomy, Washington State University, Pullman, WA 99164, USA}
\newcommand{\WSUId}{{11}}

\begin{document}

\title[Simulating binary black hole mergers using discontinuous Galerkin
methods]{Simulating binary black hole mergers using discontinuous Galerkin
  methods}

\author{
  Geoffrey Lovelace$^{\FullertonId\star}$\orcidlink{0000-0002-7084-1070},
  Kyle C.~Nelli$^{\CaltechId\star}$\orcidlink{0000-0003-2426-8768}, 
  Nils Deppe$^{\CornellPhysicsId,\CornellLeppId,\CornellCcapsId}$\orcidlink{0000-0003-4557-4115},
  Nils L.~Vu$^\CaltechId$\orcidlink{0000-0002-5767-3949},
  William Throwe$^\CornellCcapsId$\orcidlink{0000-0001-5059-4378},
  Marceline S.~Bonilla$^\FullertonId$\orcidlink{0000-0003-4502-528X},
  Alexander Carpenter$^\FullertonId$\orcidlink{0000-0002-9183-8006}, 
  Lawrence E.~Kidder$^\CornellCcapsId$\orcidlink{0000-0001-5392-7342},
  Alexandra Macedo$^\FullertonId$\orcidlink{0009-0001-7671-6377},
  Mark A.~Scheel$^\CaltechId$\orcidlink{0000-0001-6656-9134},
  Azer Afram$^\FullertonId$\orcidlink{0009-0003-2340-4059},
  Michael Boyle$^{\CornellCcapsId}$\orcidlink{0000-0002-5075-5116},
  Andrea Ceja$^{\FullertonId,\NorthwesternId}$\orcidlink{0000-0002-1681-7299},
  Matthew Giesler$^\CornellCcapsId$\orcidlink{0000-0003-2300-893X},
  Sarah Habib$^\CaltechId$\orcidlink{0000-0002-4725-4978},
  Ken Z.~Jones$^\FullertonId$\orcidlink{0009-0003-1034-0498},
  Prayush Kumar$^\IctsId$\orcidlink{0000-0001-5523-4603},
  Guillermo Lara$^\AeiId$\orcidlink{0000-0001-9461-6292},
  Denyz Melchor$^{\FullertonId,\UCLAId}$\orcidlink{0000-0002-7854-1953},
  Iago B.~Mendes$^{\CaltechId,\OberlinId}$\orcidlink{0009-0007-9845-8448},
  Keefe Mitman$^{\CornellCcapsId}$\orcidlink{0000-0003-0276-3856},
  Marlo Morales$^{\FullertonId,\WSUId}$\orcidlink{0000-0002-0593-4318},
  Jordan Moxon$^\CaltechId$\orcidlink{0000-0001-9891-8677},
  Eamonn O'Shea$^{\CornellCcapsId}$,
  Kyle Pannone$^\FullertonId$\orcidlink{0009-0005-8607-2113},
  Harald P.~Pfeiffer$^\AeiId$\orcidlink{0000-0001-9288-519X},
  Teresita Ramirez-Aguilar$^{\FullertonId,\NorthwesternId}$\orcidlink{0000-0003-0994-115X},
  Jennifer Sanchez$^{\FullertonId,\NorthwesternId}$\orcidlink{0000-0002-5335-4924},
  Daniel Tellez$^\FullertonId$\orcidlink{0009-0008-7784-2528},
  Saul A.~Teukolsky$^{\CornellCcapsId,\CaltechId}$\orcidlink{0000-0001-9765-4526},
  Nikolas A.~Wittek$^\AeiId$\orcidlink{0000-0001-8575-5450}}

\address{$^\FullertonId$\Fullerton}
\address{$^\CaltechId$\Caltech}
\address{$^\CornellPhysicsId$\CornellPhysics}
\address{$^\CornellLeppId$\CornellLepp}
\address{$^\CornellCcapsId$\CornellCcaps}
\address{$^\NorthwesternId$\Northwestern}
\address{$^\IctsId$\Icts}
\address{$^\AeiId$\Aei}
\address{$^\UCLAId$\UCLA}
\address{$^\OberlinId$\Oberlin}
\address{$^\WSUId$\WSU}
\address{$^{\star}$The first two authors contributed equally.}
\ead{glovelace@fullerton.edu}
\ead{knelli@caltech.edu}

\begin{abstract}

  Binary black holes are the most abundant source of gravitational-wave
  observations. Gravitational-wave observatories in the next decade will require
  tremendous increases in the accuracy of numerical waveforms modeling binary
  black holes, compared to today's state of the art. One approach to achieving
  the required accuracy is using spectral-type methods that scale to many
  processors. Using the
  \texttt{SpECTRE} numerical-relativity code, we present the first simulations
  of a binary black hole inspiral, merger, and ringdown using discontinuous
  Galerkin methods. The efficiency of discontinuous Galerkin methods allows us
  to evolve the binary through $\sim18$ orbits at reasonable computational cost.
  We then use \texttt{SpECTRE}'s Cauchy Characteristic Evolution (CCE) code to
  extract the gravitational waves at future null infinity.  The open-source
  nature of \texttt{SpECTRE} means this is the first time a spectral-type method
  for simulating binary black hole evolutions is available to the entire
  numerical-relativity community.

\end{abstract}


\submitto{\CQG}

\section{Introduction\label{sec:introduction}}

Binary black holes are the most abundant source of gravitational-wave
observations to date~\cite{KAGRA:2021vkt}. Realizing the scientific potential of
these observations requires accurate models of the emitted gravitational waves
as the black holes inspiral, merge, and ring down to a final, stationary
state. Building these models requires numerical-relativity (NR) simulations of binary
black holes, because analytic approximations (e.g.~the
post-Newtonian~\cite{Blanchet:2024mnz} approximation) alone break down near the
time of merger.

Since the first breakthrough simulations~\cite{Pretorius:2005gq,
  Campanelli:2005dd, Baker:2005vv}, the NR community has developed codes capable
  of evolving two black holes through inspiral, merger, and ringdown
  (see~\response{\cite{Pfeiffer:2012pc, Duez:2018jaf}} for a review). Several groups have
  used NR codes to build catalogs of gravitational waveforms for applications to
  gravitational-wave astronomy~\response{\cite{Mroue:2013xna, Jani:2016wkt, Healy:2017psd,
  Healy:2019jyf, Boyle:2019kee, Healy:2020vre, Healy:2022wdn, Ferguson:2023vta}}. 
  Today's NR codes are sufficiently accurate for the
  observations that LIGO and Virgo are making. However, observatories planned
  for the next decade, including the Einstein
  Telescope~\cite{2010CQGra..27s4002P} and Cosmic Explorer~\cite{Evans:2021gyd}
  on Earth and the Laser Interferometer Space Antenna
  (LISA)~\cite{2017arXiv170200786A} in space, will be so sensitive that they
  will require NR waveforms with a substantial increase in
  accuracy~\cite{Purrer:2019jcp, Ferguson:2020xnm, Jan:2023raq}.

Spectral-type methods are extremely efficient; this makes them a promising
avenue toward the ultimate goal of achieving the needed accuracy for future
gravitational-wave observatories. In comparison, almost all current NR codes for
evolving binary black holes use finite-difference methods, with numerical errors
decreasing as a power law with increasing resolution. However, recent results
from the \texttt{AthenaK} code~\cite{Zhu:2024utz} show that finite-difference
methods using graphics processing units (GPUs) might be another approach to
achieving the required accuracy. The Spectral Einstein Code
(\texttt{SpEC})~\cite{SpECwebsite} uses a pseudospectral method
(see~\cite{Hesthaven2007} for a review of these methods) to construct and evolve
binary-black-hole initial data. With pseudospectral methods, errors decrease
exponentially with increasing number of grids points in the computational
domain's elements (``$p$-refinement''). \texttt{SpEC}'s exponential convergence
makes it highly efficient, but its performance, and therefore the achievable
accuracy, is limited by aspects of its design. For instance, because it uses
computational domains divided into a small number of high-resolution elements,
\texttt{SpEC} simulations of binary black holes cannot scale beyond
$\mathcal{O}(10^2)$ CPU cores. \texttt{SpEC} is also a closed-source code,
unavailable to most of the NR community. Other examples of pseudospectral or
spectral methods being used for solving the initial value problem are
\texttt{Elliptica}~\cite{Rashti:2021ihv},
\texttt{FUKA}~\cite{Papenfort:2021hod}, and \texttt{bamps}~\cite{Ruter:2017iph}.
In terms of evolving spacetimes, the \texttt{Nmesh}~\cite{Tichy:2022hpa} code
has been used to successfully simulate single black holes using discontinuous
Galerkin methods, and the \texttt{bamps}~\cite{Hilditch:2015aba,
Bugner:2015gqa} code uses pseudospectral methods to evolve spacetimes with
single dynamical black holes with a focus on critical
behavior~\response{\cite{Hilditch:2017dnw, SuarezFernandez:2020wqv, Bhattacharyya:2021dti,
Fernandez:2022hyx, Renkhoff:2023nfw, Baumgarte:2023tdh, Cors:2023ncc,
Marouda:2024epb}} but has also simulated boson stars~\cite{Atteneder:2023pge}.
Recently,~\cite{Dumbser:2024tvm} performed a 0.5 orbit grazing collision of two
black holes (a similar setup to \cite{Alcubierre:2000ke}) using a finite volume
grid in the strong field region and a discontinuous-Galerkin method in the wave 
zone.

We present the first simulations of a binary black hole inspiral, merger, and
ringdown using a discontinuous Galerkin (DG) method~\cite{reed1973triangular}
(see~\cite{Hesthaven2008} for a review of DG). The efficiency of DG methods
allows us to evolve the binary through $\sim18$ orbits at reasonable
computational cost:  DG, being a spectral-type method, has exponential
convergence with $p$-refinement. \response{For context, 18 orbits is slightly less than
the median (20) for binary-black-hole simulations in the SXS
catalog~\cite{Boyle:2019kee} (which also uses a spectral-type method) but larger
than almost all of the simulations in the RIT and Maya
catalogs~\cite{Healy:2022wdn, Ferguson:2023vta} (which use finite-difference
methods). We chose the length of \texttt{SpECTRE}'s first binary-black-hole simulation
largely out of convenience, balancing a desire to demonstrate \texttt{SpECTRE}'s
capability with minimizing turnaround time as we tested and fine-tuned our
methods. We expect that using \texttt{SpECTRE} to simulate more orbits would be
straightforward, without requiring changes to the code, although extending the
length beyond 100 orbits would likely require implementing in \texttt{SpECTRE} similar
techniques as those discussed in~\cite{Szilagyi:2015rwa}, which presents a
175-orbit \texttt{SpEC} binary-black-hole simulation.}

Specifically, in this work, we
present \texttt{SpECTRE}'s~\cite{spectrecode_2024} first simulations of $\sim18$
orbits of inspiral, merger, and ringdown of an equal-mass, non-spinning binary
black hole, using DG methods. We then use \texttt{SpECTRE}'s Cauchy
Characteristic Evolution module~\response{\cite{2020PhRvD.102d4052M, Moxon:2021gbv,
spectrecode_2024}} to evolve the gravitational waves to future null infinity.
These results demonstrate that DG methods can provide high-accuracy
gravitational waveforms from binary black hole mergers for application to
gravitational-wave data analysis. By implementing our approach in
\texttt{SpECTRE}, an open-source NR code, we are also making a
spectral-type binary-black-hole evolution code available to the entire
NR community for the first time.

The rest of this paper is organized as follows. In \S\ref{sec:methods}, we
discuss the numerical methods used in \texttt{SpECTRE}'s binary-black-hole
simulations. Then, in \S\ref{sec:results}, we first test our method's stability
with simulations of single black holes before presenting results for simulations
of binary black holes with \texttt{SpECTRE}. We briefly conclude in
\S\ref{sec:conclusion}.

\section{Methods\label{sec:methods}}

\subsection{Equations of Motion}
\label{sec:equations}

We adopt the standard 3+1 form of the spacetime metric,~(see,
e.g.,~\cite{Baumgarte:2010ndz, 2013rehy.book.....R}),
\begin{eqnarray}
  \label{eq:spacetime metric}
  ds^2 &= g_{ab}dx^a dx^b =-\alpha^2 dt^2 + \gamma_{ij}
         \left(dx^i+\beta^i dt\right) \left(dx^j +\beta^j dt\right),
\end{eqnarray}
where $\alpha$ is the lapse, $\beta^i$ the shift vector, and $\gamma_{ij}$ is
the spatial metric.  We use the Einstein summation convention, summing over
repeated indices.  Latin indices from the first part of the alphabet
$a,b,c,\ldots$ denote spacetime indices ranging from $0$ to $3$, while Latin
indices $i,j,\ldots$ are purely spatial, ranging from $1$ to $3$. We work in
units where $c = G = 1$.

We evolve the first-order generalized harmonic (FOGH) system, given
by~\cite{Lindblom:2005qh},
\begin{eqnarray}
  \label{eq:fosh gh metric evolution}
  \partial_t g_{ab}
  &=\left(1+\gamma_1\right)\beta^k\partial_k g_{ab}
    -\alpha \Pi_{ab}-\gamma_1\beta^i\Phi_{iab}
    +\gamma_1v^k_g\left(\partial_k g_{ab}-\Phi_{kab}\right), \\
  \label{eq:fosh gh metric derivative evolution}
  \partial_t\Phi_{iab}
  &=\beta^k\partial_k\Phi_{iab} - \alpha \partial_i\Pi_{ab}
    + \alpha \gamma_2\partial_ig_{ab}
    +\frac{1}{2}\alpha n^c n^d\Phi_{icd}\Pi_{ab} \nonumber \\
  &+ \alpha \gamma^{jk}n^c\Phi_{ijc}\Phi_{kab}
    -\alpha \gamma_2\Phi_{iab},\\
  \label{eq:fosh gh metric conjugate evolution}
  \partial_t\Pi_{ab}
  &=\beta^k\partial_k\Pi_{ab} - \alpha \gamma^{ki}\partial_k\Phi_{iab}
    + \gamma_1\gamma_2\beta^k\partial_kg_{ab} \nonumber \\
  &+2\alpha g^{cd}\left(\gamma^{ij}\Phi_{ica}\Phi_{jdb}
    - \Pi_{ca}\Pi_{db} - g^{ef}\Gamma_{ace}\Gamma_{bdf}\right) \nonumber \\
  &-2\alpha \nabla_{(a}H_{b)}
    - \frac{1}{2}\alpha n^c n^d\Pi_{cd}\Pi_{ab}
    - \alpha n^c \Pi_{ci}\gamma^{ij}\Phi_{jab} \nonumber \\
  &+\alpha \gamma_0\left(2\delta^c{}_{(a} n_{b)}
    - g_{ab}n^c\right)\mathcal{C}_c -\gamma_1\gamma_2 \beta^i\Phi_{iab},
\end{eqnarray}
where $g_{ab}$ is the spacetime metric, $\Phi_{iab}=\partial_i g_{ab}$,
$\Pi_{ab} = n^c\partial_cg_{ab}$, $n^a$ is the unit normal vector to the spatial
slice, $\gamma_0$ damps the 1-index or gauge constraint
$\mathcal{C}_a=H_a+\Gamma_a$, $\gamma_1$ controls the linear degeneracy of the
system, $\gamma_2$ damps the 3-index constraint
$\mathcal{C}_{iab}=\partial_i g_{ab}-\Phi_{iab}$, $\Gamma_{abc}$ are the
spacetime Christoffel symbols of the first kind, $\Gamma_a=g^{bc}\Gamma_{bca}$,
and $v^k_g$ is the grid/mesh velocity as discussed in \S\ref{sec:dg method}.

The gauge source function $H_a$ can be any arbitrary function depending only
upon the spacetime coordinates $x^a$ and $g_{ab}$, but not derivatives of
$g_{ab}$, since that may spoil the strong hyperbolicity of the
system~\cite{Lindblom:2007xw, Szilagyi:2009qz}.

Defining $s_i$ to be the unit normal covector to a 2d surface with
$s_a=(0,s_i)$, and $s^a=g^{ab}s_b$, the characteristic fields for the FOGH
system are~\cite{Lindblom:2005qh}
\begin{eqnarray}
  \label{eq:fosh gh w g}
  w^{g}_{ab} &= g_{ab}, \\
  \label{eq:fosh gh w 0}
  w^{0}_{iab} &= (\delta^k_i-s^k s_i)\Phi_{kab}, \\
  \label{eq:fosh gh w pm}
  w^{\pm}_{ab} &= \Pi_{ab}\pm s^i\Phi_{iab} -\gamma_2 g_{ab},
\end{eqnarray}
with associated characteristic speeds
\begin{eqnarray}
  \label{eq:fosh gh char speed g}
  \lambda_{w^g} =& -(1+\gamma_1)\beta^i s_i -(1+\gamma_1)v^i_g s_i, \\
  \label{eq:fosh gh char speed 0}
  \lambda_{w^0} =& -\beta^i s_i -v^i_g s_i, \\
  \label{eq:fosh gh char speed pm}
  \lambda_{w^\pm} =& \pm \alpha - \beta^i s_i - v^i_g s_i,
\end{eqnarray}
where we denote the velocity of the grid/mesh as $v^i_g$ (see~\S\ref{sec:dg
  method} for details on our moving mesh method). The evolved variables as a
function of the characteristic fields are given by
\begin{eqnarray}
  \label{eq:fosh gh metric from char}
  g_{ab} &= w^g_{ab}, \\
  \label{eq:fosh gh pi from char}
  \Pi_{ab} &= \frac{1}{2}\left(w^+_{ab} + w^-_{ab}\right)+\gamma_2 w^g_{ab}, \\
  \label{eq:fosh gh phi from char}
  \Phi_{iab} &= \frac{1}{2}\left(w^+_{ab} - w^-_{ab}\right)s_i + w^0_{iab}.
\end{eqnarray}

The constraints for the FOGH system are~\cite{Lindblom:2005qh}
\begin{eqnarray}
  \label{eq:gauge constraint}
  \mathcal{C}_a
  &=H_a+\Gamma_a, \\
  \label{eq:2 index constraint}
  \mathcal{C}_{ia}
  &= \gamma^{jk}\partial_j \Phi_{ika}
    - \frac{1}{2} \gamma_a^jg^{cd}\partial_j \Phi_{icd}
    + n^b \partial_i \Pi_{ba}
    - \frac{1}{2} n_a g^{cd}\partial_i\Pi_{cd}\nonumber\\
  &+ \partial_i H_a + \frac{1}{2} \gamma_a^j \Phi_{jcd} \Phi_{ief} g^{ce}g^{df}
    + \frac{1}{2} \gamma^{jk} \Phi_{jcd} \Phi_{ike} g^{cd}n^e n_a \nonumber\\
  &- \gamma^{jk}\gamma^{mn}\Phi_{jma}\Phi_{ikn}
    + \frac{1}{2} \Phi_{icd} \Pi_{be} n_a \left(g^{cb}g^{de}
    + \frac{1}{2}g^{be} n^c n^d\right) \nonumber\\
  &- \Phi_{icd} \Pi_{ba} n^c \left(g^{bd} +\frac{1}{2} n^b n^d\right)
    + \frac{1}{2} \gamma_2 \left(n_a g^{cd}
    - 2 \delta^c_a n^d\right) \mathcal{C}_{icd}, \\
  \label{eq:3 index constraint}
  \mathcal{C}_{iab}
  &=\partial_i g_{ab}-\Phi_{iab}, \\
  \label{eq:4 index constraint}
  \mathcal{C}_{ijab}
  &=\partial_i\Phi_{jab}-\partial_j\Phi_{iab},
\end{eqnarray}
and
\begin{eqnarray}
  \label{eq:F constraint}
  \mathcal{F}_{a}
  &=
   \frac{1}{2} \gamma_a^i g^{bc}\partial_i \Pi_{bc}
   - \gamma^{ij} \partial_i \Pi_{ja}
   - \gamma^{ij} n^b \partial_i \Phi_{jba}
   + \frac{1}{2} n_a g^{bc} \gamma^{ij} \partial_i \Phi_{jbc}
   \nonumber \\ &
   + n_a \gamma^{ij} \partial_i H_j
   + \gamma_a^i \Phi_{ijb} \gamma^{jk}\Phi_{kcd} g^{bd} n^c
   - \frac{1}{2} \gamma_a^i \Phi_{ijb} \gamma^{jk}
     \Phi_{kcd} g^{cd} n^b
   \nonumber \\ &
   - \gamma_a^i n^b \partial_i H_b
   + \gamma^{ij} \Phi_{icd} \Phi_{jba} g^{bc} n^d
   - \frac{1}{2} n_a \gamma^{ij} \gamma^{mn} \Phi_{imc} \Phi_{njd}g^{cd}
   \nonumber \\ &
   - \frac{1}{4}  n_a \gamma^{ij}\Phi_{icd}\Phi_{jbe}
      g^{cb}g^{de}
   + \frac{1}{4}  n_a \Pi_{cd} \Pi_{be}
      g^{cb}g^{de}
   - \gamma^{ij} H_i \Pi_{ja}
   \nonumber \\ &
   - n^b \gamma^{ij} \Pi_{b i} \Pi_{ja}
   - \frac{1}{4}  \gamma_a^i \Phi_{icd} n^c n^d \Pi_{be}
     g^{be}
   + \frac{1}{2} n_a \Pi_{cd} \Pi_{be}g^{ce}
     n^d n^b
   \nonumber \\ &
   + \gamma_a^i \Phi_{icd} \Pi_{be} n^c n^b g^{de}
   - \gamma^{ij}\Phi_{iba} n^b \Pi_{je} n^e
   - \frac{1}{2} \gamma^{ij}\Phi_{icd} n^c n^d \Pi_{ja}
   \nonumber \\ &
   - \gamma^{ij} H_i \Phi_{jba} n^b
   + \gamma_{a}^i \Phi_{icd} H_b g^{bc} n^d
   +\gamma_2\left(\gamma^{id}\mathcal{C}_{ida}
   -\frac{1}{2}  \gamma_a^ig^{cd}\mathcal{C}_{icd}\right)
   \nonumber \\ &
   + \frac{1}{2} n_a \Pi_{cd}g^{cd} H_b n^b
   - n_a \gamma^{ij} \Phi_{ijc} H_d g^{cd}
   +\frac{1}{2}  n_a \gamma^{ij} H_i \Phi_{jcd}g^{cd}.
\end{eqnarray}
While only the gauge constraint~(\ref{eq:gauge constraint}) and 3-index
constraint~(\ref{eq:3 index constraint}) are damped, all constraints can be
monitored to check the accuracy of the numerical simulation. All the constraints
can be combined into a scalar, the constraint energy, given
by~\cite{Lindblom:2005qh}
\begin{eqnarray}
  \label{eq:constraint energy}
  \mathcal{E}
  & = \delta^{ab}\left[ \mathcal{C}_a \mathcal{C}_b
    + \left(\mathcal{F}_a \mathcal{F}_b
    + \mathcal{C}_{ia} \mathcal{C}_{jb} \gamma^{ij}\right)\right]
    \nonumber \\
  &+ \delta^{ab}\delta^{cd}\left(\mathcal{C}_{iac} \mathcal{C}_{jbd}
    \gamma^{ij}
    + \mathcal{C}_{ikac} \mathcal{C}_{jlbd}\gamma^{ij} \gamma^{kl}\right).
\end{eqnarray}
In practice we have also found that it is typically only necessary to monitor
violations of the constraints $\mathcal{C}_a$ and $\mathcal{C}_{iab}$, because
they typically grow first and the other violations grow as a consequence.

\subsection{Discontinuous Galerkin \response{(DG)} Method\label{sec:dg method}}

\texttt{SpECTRE} uses a \response{DG} method for the spatial
discretization. We refer readers to \cite{Teukolsky:2015ega, Deppe:2021ada}
and references therein for a detailed discussion of the method and its
implementation in \texttt{SpECTRE}; here we summarize the necessary
results. The FOGH equations are a first-order strongly hyperbolic system in
non-conservative form, which takes the general form
\begin{eqnarray}
  \label{eq:fosh system}
  \partial_t u_\alpha + B_{\alpha\beta}^i(u_\alpha) \partial_i u_{\beta} =
  S_\alpha(u_\alpha),
\end{eqnarray}
where $u_\alpha=\{g_{ab},\Phi_{iab},\Pi_{ab}\}$ is the state vector of evolved
variables, $B_{\alpha\beta}^i(u)$ is a matrix that depends only on $u_\alpha$,
and $S_\alpha(u)$ are source terms that also only depend on $u_\alpha$. We
denote the logical coordinates of our Legendre-Gauss-Lobatto DG scheme by
$\{\hat{t}, \xi^{\hat{\imath}}\}=\{\hat{t},\xi,\eta,\zeta\}$ and the inertial
coordinates as $\{t=\hat{t}, x^i(\hat{t}, \xi^{\hat{\imath}})\}$. We are using a
moving mesh as in~\cite{Scheel:2006gg, Teukolsky:2015ega}; therefore, the
mapping from logical to inertial coordinates is time dependent. We denote the
determinant of the spatial Jacobian of this map as
\begin{eqnarray}
  \label{eq:2}
  J=\det\left(\frac{\partial x^i}{\partial \xi^{\hat{\imath}}}\right),
\end{eqnarray}
and the grid or mesh velocity by~\cite{Scheel:2006gg, Teukolsky:2015ega}
\begin{eqnarray}
  \label{eq:grid velocity}
  v_g^i=\partial_{\hat{t}} x^i.
\end{eqnarray}

\begin{figure}
  \centering
  \includegraphics[width=0.5\linewidth]{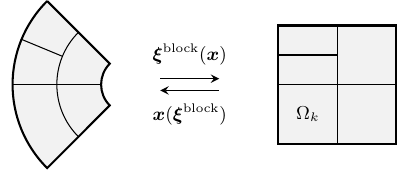}
\caption{A diagram of the forward and reverse mappings $x^i(\hat{t},
\xi^{\hat{\imath}})$ and $\xi^{\hat{\imath}}(t,x^i)$, respectively, from logical
(right side of the diagram) to inertial coordinates (left side of the diagram)
for an element $\Omega_k$. \label{fig:logical_map}}
\end{figure}

\texttt{SpECTRE} decomposes the domain into a set of non-overlapping hexahedra
which are deformed using the map $x^i(\hat{t}, \xi^{\hat{\imath}})$ illustrated
in figure~\ref{fig:logical_map}. \texttt{SpECTRE} uses a different number of
grid points in each logical direction, which we denote by $N_{\breve{\imath}}$
in the $\xi$ direction, $N_{\breve{\jmath}}$ in the $\eta$ direction, and
$N_{\breve{k}}$ in the $\zeta$ direction below.

Since we use a moving mesh, we evolve the system
\begin{eqnarray}
  \label{eq:fosh system moving mesh}
  \partial_{\hat{t}} u_\alpha + \left[B_{\alpha\beta}^i(u_\alpha) -
  v^i_g\delta_{\alpha\beta}\right] \partial_i u_\beta = S(u_\alpha).
\end{eqnarray}
We denote grid point and modal indices with a breve,
i.e.~$u_{\breve{\imath}\breve{\jmath}\breve{k}}$ is the value of $u$ at the
grid point $(\breve{\imath},\breve{\jmath},\breve{k})$. The semi-discrete
equations are given by~\cite{Scheel:2006gg, Teukolsky:2015ega}
\begin{eqnarray}
  \label{eq:dg strong form semi-discrete}
  (\partial_{\hat{t}} u_\alpha)_{\breve{\imath}\breve{\jmath}\breve{k}}
  &=(S_\alpha)_{\breve{\imath}\breve{\jmath}\breve{k}}
    -\left(B_{\alpha\beta}^i-
    v^i_g\delta_{\alpha\beta}\right)_{\breve{\imath}\breve{\jmath}\breve{k}}
    \left[\left(\frac{\partial \xi^{\hat{1}}}{\partial
    x^i}\right)_{\breve{\imath}\breve{\jmath}\breve{k}}
    \sum_{\breve{l}} D^{(\hat{1})}_{\breve{\imath}\breve{l}}
    (u_\beta)_{\breve{l}\breve{\jmath}\breve{k}}\right.
    \nonumber \\
  &-\left.\left(\frac{\partial \xi^{\hat{2}}}{\partial
    x^i}\right)_{\breve{\imath}\breve{\jmath}\breve{k}}
    \sum_{\breve{l}} D^{(\hat{2})}_{\breve{\jmath}\breve{l}}
    (u_\beta)_{\breve{\imath}\breve{l}\breve{k}}
    +\left(\frac{\partial \xi^{\hat{3}}}{\partial
    x^i}\right)_{\breve{\imath}\breve{\jmath}\breve{k}}
    \sum_{\breve{l}} D^{(\hat{3})}_{\breve{k}\breve{l}}
    (u_\beta)_{\breve{\imath}\breve{\jmath}\breve{l}}\right]
    \nonumber \\
  &- \frac{\delta_{N_{\breve{k}}\breve{k}}}
    {w_{\breve{k}}J_{\breve{\imath}\breve{\jmath}\breve{k}}}
    \left\{
    \left[J\sqrt{
    \frac{\partial\xi^{\hat{3}}}{\partial x^i} \gamma^{ij}
    \frac{\partial\xi^{\hat{3}}}{\partial x^j}}
    D_\alpha\right]_{\breve{\imath}\breve{\jmath}N_{\breve{k}}}
    +\left[J\sqrt{
    \frac{\partial\xi^{\hat{3}}}{\partial x^i} \gamma^{ij}
    \frac{\partial\xi^{\hat{3}}}{\partial x^j}} D_\alpha
    \right]_{\breve{\imath}\breve{\jmath}0}\right\}
    \nonumber \\
  &- \frac{\delta_{N_{\breve{\jmath}}\breve{\jmath}}}
    {w_{\breve{\jmath}}J_{\breve{\imath}\breve{\jmath}\breve{k}}}
    \left\{
    \left[J\sqrt{
    \frac{\partial\xi^{\hat{3}}}{\partial x^i} \gamma^{ij}
    \frac{\partial\xi^{\hat{3}}}{\partial x^j}}
    D_\alpha\right]_{\breve{\imath}N_{\breve{\jmath}}\breve{k}}
    +\left[J\sqrt{
    \frac{\partial\xi^{\hat{3}}}{\partial x^i} \gamma^{ij}
    \frac{\partial\xi^{\hat{3}}}{\partial x^j}} D_\alpha
    \right]_{\breve{\imath}0\breve{k}}\right\}
    \nonumber \\
  &- \frac{\delta_{N_{\breve{\imath}}\breve{\imath}}}
    {w_{\breve{\imath}}J_{\breve{\imath}\breve{\jmath}\breve{k}}}
    \left\{
    \left[J\sqrt{
    \frac{\partial\xi^{\hat{3}}}{\partial x^i} \gamma^{ij}
    \frac{\partial\xi^{\hat{3}}}{\partial x^j}}
    D_\alpha\right]_{N_{\breve{\imath}}\breve{\jmath}\breve{k}}
    +\left[J\sqrt{
    \frac{\partial\xi^{\hat{3}}}{\partial x^i} \gamma^{ij}
    \frac{\partial\xi^{\hat{3}}}{\partial x^j}} D_\alpha
    \right]_{0\breve{\jmath}\breve{k}}\right\},
\end{eqnarray}
where $w_{\breve{\imath}}$ are the Legendre-Gauss-Lobatto integration weights.
We use a method of lines approach to integrate these in time, with the details
discussed in \S\ref{sec:time integration} below.

For the boundary terms $D_\alpha$, we use an upwind multi-penalty
method~\cite{Hesthaven1997, Hesthaven1999, Hesthaven2000, Hesthaven2007} given
by
\begin{eqnarray}
  \label{eq:upwind gh metric}
  D_{g_{ab}}
  &= \tilde{\lambda}_{w^{g}}^{\mathrm{ext}} w^{\mathrm{ext},g}_{ab}
    - \tilde{\lambda}_{w^{g}}^{\mathrm{int}} w^{\mathrm{int},g}_{ab}, \\
  \label{eq:upwind gh pi}
  D_{\Pi_{ab}}
  &= \frac{1}{2}\left(\tilde{\lambda}_{w^+}^{\mathrm{ext}} w^{\mathrm{ext},+}_{ab}
    + \tilde{\lambda}_{w^-}^{\mathrm{ext}}
    w^{\mathrm{ext},-}_{ab}\right)
    + \tilde{\lambda}_{w^g}^\mathrm{ext}\gamma_2
    w^{\mathrm{ext},g}_{ab}
    \nonumber \\
  &-\frac{1}{2}\left(\tilde{\lambda}_{w^+}^{\mathrm{int}}
    w^{\mathrm{int},+}_{ab} +
    \tilde{\lambda}_{w^-}^{\mathrm{int}}
    w^{\mathrm{int},-}_{ab}\right)
    - \tilde{\lambda}_{w^g}^\mathrm{int}\gamma_2
    w^{\mathrm{int},g}_{ab} , \\
  \label{eq:upwind gh phi}
  D_{\Phi_{iab}}
  &= \frac{1}{2}\left(\tilde{\lambda}_{w^+}^{\mathrm{ext}}
    w^{\mathrm{ext},+}_{ab}
    - \tilde{\lambda}_{w^-}^{\mathrm{ext}}
    w^{\mathrm{ext},-}_{ab}\right)s_i^{\mathrm{ext}}
    + \tilde{\lambda}_{w^0}^{\mathrm{ext}}
    w^{\mathrm{ext},0}_{iab}
    \nonumber \\
  &-\frac{1}{2}\left(\tilde{\lambda}_{w^+}^{\mathrm{int}}
    w^{\mathrm{int},+}_{ab}
    - \tilde{\lambda}_{w^-}^{\mathrm{int}}
    w^{\mathrm{int},-}_{ab}\right)s_i^{\mathrm{int}}
    - \tilde{\lambda}_{w^0}^{\mathrm{int}}
    w^{\mathrm{int},0}_{iab},
\end{eqnarray}
where the spatial normal vector to the element interface $s^{\mathrm{int}}_i$ is
pointing out of the DG element and $\tilde{\lambda}=0$ if $\lambda>0$, otherwise
$\tilde{\lambda}=\lambda$, i.e.~$\tilde{\lambda}=\lambda\Theta(-\lambda)$.
Note that we assume $s^{\mathrm{ext}}_i$ and $s^{\mathrm{int}}_i$ point in the
same direction. Also note that these boundary flux terms differ from the
multi-penalty approach used in \texttt{SpEC} by a factor of $2$. That is,
\begin{eqnarray}
  \label{eq:boundary term compared to spec}
  D^{\mathtt{SpECTRE}} = 2 D^{\mathtt{SpEC}},
\end{eqnarray}
ultimately because the lifting terms are different. In \texttt{SpEC} and,
similarly in \texttt{bamps}\cite{Hilditch:2015aba}, the penalty term is derived
from requiring that the total energy be non-increasing, while in
\texttt{SpECTRE} the terms come from an integration by parts when deriving the
semi-discrete DG equations.

\subsection{Boundary conditions\label{sec:boundary conditions}}

At the outer radial boundary, we apply constraint-preserving boundary
conditions~\cite{Lindblom:2005qh, Rinne:2007ui} by adding terms to the time
derivative of the characteristic fields and thus also the time derivatives of
the evolved variables. We use the characteristic fields and speeds defined
in~\S\ref{sec:equations}. We define $d_{\hat{t}} w_{\hat{\alpha}}$ as the time
derivatives substituted into the transformation equations to the characteristic
fields. That is,
\begin{eqnarray}
  \label{eq:fosh gh dtw g}
  d_{\hat{t}}w^{g}_{ab} &= \partial_{\hat{t}} g_{ab}, \\
  \label{eq:fosh gh dtw 0}
  d_{\hat{t}} w^{0}_{iab} &= (\delta^k_i-s^k s_i)\partial_{\hat{t}} \Phi_{kab},
  \\
  \label{eq:fosh gh dtw pm}
  d_{\hat{t}} w^{\pm}_{ab} &= \partial_{\hat{t}} \Pi_{ab}\pm
                             s^i\partial_{\hat{t}} \Phi_{iab}
                             -\gamma_2\partial_{\hat{t}} g_{ab}.
\end{eqnarray}
We also define $D_{\hat{t}} w_{\hat{\alpha}}$ as the characteristic field
transformation of the volume right-hand-side, i.e. $\partial_{\hat{t}} u_\alpha$
\textit{without} any boundary terms. Finally, for brevity we define the
projection tensor $P_{ab}=g_{ab}+n_a n_b-s_as_b$, the inward directed null
vector field $k^a=(n^a-s^a)/\sqrt{2}$, and the outgoing null vector field
$l^a=(n^a+s^a)/\sqrt{2}$.

The fields $d_{\hat{t}} w^g_{ab}$ and $d_{\hat{t}} w^0_{iab}$ are determined
solely by the constraint-preserving boundary condition, while the boundary
condition for $d_{\hat{t}} w^{-}_{ab}$ is composed of three parts: the
constraint preserving part, the physical part, and the gauge part. We denote
these as $B_{ab}^{C}$, $B_{ab}^P$ and $B_{ab}^G$. With this, the boundary
conditions imposed on the fields are
\begin{eqnarray}
  \label{eq:fosh gh bc d_t char}
  d_{\hat{t}}w^{g}_{ab} &= D_{\hat{t}} w^g_{ab}
                          +\lambda_{w^g}s^i\mathcal{C}_{iab}, \\
  d_{\hat{t}}w^{0}_{kab} &= D_{\hat{t}} w^0_{ab}
                           +\lambda_{w^0}s^iP^j{}_k\mathcal{C}_{ijab}, \\
  d_{\hat{t}}w^{-}_{ab} &= D_{\hat{t}} w^-_{ab}
                          +\lambda_{w^-}\left[B_{ab}^C+B_{ab}^P+B_{ab}^G\right].
\end{eqnarray}
Transforming to the evolved variables we find that the following terms need to
be added in order to impose the boundary condition,
\begin{eqnarray}
  \label{eq:fosh gh constraing preserving bc g}
  \partial_{\hat{t}} g_{ab}&\to \partial_{\hat{t}} g_{ab} +
                             \lambda_{w^g}s^i\mathcal{C}_{iab}, \\
  \label{eq:fosh gh constraing preserving bc pi}
  \partial_{\hat{t}} \Pi_{ab}&\to \partial_{\hat{t}} \Pi_{ab}
                       +\frac{1}{2}\lambda_{w^-}
                       \left[B_{ab}^C+B_{ab}^P+B_{ab}^G\right]
                       +\gamma_2 \lambda_{w^g}s^i\mathcal{C}_{iab}, \\
  \label{eq:fosh gh constraing preserving bc phi}
  \partial_{\hat{t}} \Phi_{iab}&\to\partial_{\hat{t}} \Phi_{iab}
                         -\frac{s_i}{2}\lambda_{w^-}
                         \left[B_{ab}^C+B_{ab}^P+B_{ab}^G\right]
                         +\lambda_{w^0}s^iP^j{}_k\mathcal{C}_{ijab}.
\end{eqnarray}

We now need to specify the $B_{ab}$ boundary conditions. The
constraint-preserving part is
\begin{eqnarray}
  \label{eq:fosh gh bc constraint preserving}
  B_{ab}^C=\sqrt{2}\left(\frac{1}{2}P_{ab}l^c + \frac{1}{2}l_al_bk^c -
  l_{(a}P_{b)}{}^c\right)\left(\mathcal{F}_c-s^k\mathcal{C}_{kc}\right).
\end{eqnarray}
The physical boundary conditions are determined by the propagating parts of the
Weyl curvature tensor. That is,
\begin{eqnarray}
  \label{eq:fosh gh bc physical}
  B_{ab}^P=\left(P_a{}^cP_b{}^d-\frac{1}{2}P_{ab}P^{cd}\right)\left[C^-_{cd}
  -\gamma_2s^i\mathcal{C}_{icd}\right],
\end{eqnarray}
where $C^-_{ab}$ is the inward propagating part of the Weyl tensor, given by
\begin{eqnarray}
  \label{eq:fosh gh weyl propagating}
  C^{\pm}_{ab}=\left(P_a{}^cP_b{}^d-\frac{1}{2}P_{ab}P^{cd}\right)\left(n^e\mp
  s^e\right)\left(n^f\mp s^f\right)C_{cedf}.
\end{eqnarray}
For the simulations presented here, we set $C^-_{ab}=0$, though
Cauchy-Characteristic matching~\cite{Ma:2023qjn} can be used to prescribe a more
physically motivated boundary condition. Recently
\cite{Buchman:2024zsb} presented an alternative approach to
Cauchy-Characteristic matching for providing high-order non-reflecting boundary
conditions. Finally, the gauge boundary condition is set using a Sommerfeld
condition on the components not set by the constraint-preserving and physical
boundary conditions. The projector for the gauge boundary condition is given by
\begin{eqnarray}
  \label{eq:fosh gh bc gauge projection}
  \delta_a^c\delta_b^d-P^{C}_{ab}{}^{cd}-P^P_{ab}{}^{cd}
  &=\delta_a^c\delta_b^d-\frac{1}{2}P_{ab}P^{cd}+2l_{(a}P_{b)}{}^{(c}k^{d)}
    \nonumber \\
  &-l_al_bk^ck^d-P_a{}^cP_b{}^d+\frac{1}{2}P_{ab}P^{cd} \nonumber \\
  &=\delta_a^c\delta_b^d+2l_{(a}P_{b)}{}^{(c}k^{d)}
    -l_al_bk^ck^d-P_a{}^cP_b{}^d.
\end{eqnarray}
The Sommerfeld condition is
\begin{eqnarray}
  \label{eq:fosh gh bc gauge}
  B_{ab}^G=\frac{1}{\lambda_{w^-}}
  \left(2l_{(a}P_{b)}{}^{(c}k^{d)} - 2k_{(a}l_{b)}k^{(c}l^{d)}
  -k_ak_bl^cl^d\right)\left(\gamma_2 - \frac{1}{r}\right)\partial_t
  g_{cd}.
\end{eqnarray}

When evolving spacetimes with black holes, we excise the interior of the black
hole as is done in \texttt{SpEC}~\cite{Lindblom:2005qh}. At excision boundaries,
all information is flowing out of the grid and into the black hole, so no
boundary condition needs to be applied. However, we monitor the characteristic
speeds,~(\ref{eq:fosh gh char speed g}-\ref{eq:fosh gh char speed pm}),
and terminate the code if any of them point into the computational domain. We
denote the radius of the excision surfaces by $r_{\mathrm{exc}}$. See 
\S\ref{sec:control systems} for a brief explanation of how we control
$r_{\mathrm{exc}}$ to avoid any characteristic speed pointing into the
computational domain.

\subsection{Spectral filter\label{sec:gh spectral filter}}
We use an exponential filter applied to the spectral coefficient $c_i$ in order
to eliminate aliasing-driven instabilities. Specifically, for a 1d spectral
expansion
\begin{eqnarray}
  \label{eq:u spectral expansion}
  u(x)=\sum_{\breve{\imath}=0}^{N}c_{\breve{\imath}} P_{\breve{\imath}}(x),
\end{eqnarray}
where $P_{\breve{\imath}}(x)$ are the Legendre polynomials, we use the filter
\begin{eqnarray}
  \label{eq:exponential filter}
  c_{\breve{\imath}} \to c_{\breve{\imath}}
  \exp\left[-a\left(\frac{i}{N}\right)^{2b}\right].
\end{eqnarray}
We choose the parameters $a=64$ and $b=210$ so that only the highest
spectral mode is filtered. We apply the filter to all FOGH variables $g_{ab}$,
$\Phi_{iab}$ and $\Pi_{ab}$. Note that the filter drops the order of convergence
for the FOGH variables from $\mathcal{O}(N+1)$ to $\mathcal{O}(N)$ on the DG
grid, but is necessary for stability.

\subsection{Time integration\label{sec:time integration}}
We decompose the system using the method of lines and solve the resulting
differential equations using a local adaptive time-stepper based on the
Adams-Moulton predictor-corrector method~\cite{ThroweLTSAM}.  The
step size in each element is chosen based on an estimate of the truncation error
of the time step, using the algorithm described in \cite{NumericalRecipes}
\S17.2.1. The specific values for the absolute and relative tolerances are
given in \S\ref{sec:results}. As the time-stepping algorithm
is more efficient for aligned steps of the same size, the step size in each
element is rounded down to a value of the form $0.1 M / 2^n$ for some
non-negative integer $n$.  For the highest-resolution binary-black-hole run in
\S\ref{sec:bbh results}, 
this results in the most-demanding element taking $2^6$--$2^8$ steps for each
step on the least demanding element for most of the inspiral. At the time of
merger, this can increase to as high as $2^{11}$ steps for the most-demanding
element for each step on the least demanding element.

\subsection{Gauge condition}\label{sec:dh_gauge}
We evolve binary black holes (\S\ref{sec:bbh results}) using the Damped Harmonic
gauge condition~\cite{Szilagyi:2009qz, Deppe2018:uye}:
\begin{eqnarray}
  H_a &=& \left[\mu_{L1} \log\left(\sqrt{\gamma}/\alpha\right) + \mu_{L2}
          \log\left(1/\alpha\right)\right] n_a - \mu_S g_{ai} \beta^i / \alpha,
\end{eqnarray}
using
\begin{eqnarray}
  \mu_{L1} &=& A_{L1} e^{-(r/\sigma_r)^2}
               \left[\log(\sqrt{\gamma}/\alpha)\right]^{e_{L1}},\\
  \mu_{L2} &=& A_{L2} e^{-(r/\sigma_r)^2}
               \left[\log(\sqrt{\gamma}/\alpha)\right]^{e_{L2}},\\
  \mu_{S} &=& A_{S} e^{-(r/\sigma_r)^2}
              \left[\log(\sqrt{\gamma}/\alpha)\right]^{e_{S}},
\end{eqnarray} where $r$ is the coordinate distance from the origin. This
condition is designed to drive $\sqrt{\gamma}$ and $\alpha$ to one, while damping
out oscillations in the shift. This is because we observe an explosive growth in
$\sqrt{\gamma}$ and a rapid collapse in $\alpha$ as the black holes merge. In
practice, this ensures coordinates remain sufficiently well behaved throughout
inspiral, merger, and ringdown. The amplitudes $A_{L1}$, $A_{L2}$, and $A_{S}$ and
exponents $e_{L1}$, $e_{L2}$ and $e_{S}$ control the amount of damping, and the
spatial decay width $\sigma_r$ ensures that at large distances, the gauge reduces
to harmonic gauge (i.e., to $H_a = 0$). In this paper, we choose
$A_{L1}=A_S = 1$, $A_{L2}=0$, $e_{L1}=e_{L2}=e_{S}=2$, and $\sigma_r = 100 /
\sqrt{34.54}$. This choice for $\sigma_r$ ensures that the spatial decay
Gaussian falls to $10^{-15}$ at a distance $r=100$ from the origin.

For some of the single black-hole evolutions (\S\ref{sec:kerr black hole}), we
instead choose $H_a$ to be $\Gamma_a$ of the analytic initial data. For other
single black-hole evolutions, we evolve in harmonic gauge, setting $H_a=0$
everywhere.  For the gauge wave evolution (\S\ref{sec:gauge wave}), we set
$H_a(t, x^i)$ to the value of $\Gamma_a(t, x^i)$ of the gauge wave analytic
solution.

\subsection{Control systems\label{sec:control systems}} When evolving the FOGH
system, if there are black holes, the physical singularities inside of the black
holes must be excised from the computational domain. To position the excisions
with our moving mesh (described in \S\ref{sec:dg method}), we use a feedback
control system similar to what is presented in~\cite{Scheel:2006gg}
and~\cite{Hemberger:2012jz}. As discussed in \S\ref{sec:boundary conditions},
the excision surfaces must have all characteristic speeds pointing out of the
computational domain, so that no boundary condition must be imposed. In practice
this means that the excision surfaces must remain inside the apparent horizons,
with the caveat that having them too close to the singularity causes
instabilities. In practice the excision surfaces are kept at approximately
95-99\% of the apparent horizons' radii.

Since we \textit{a priori} do not know the motion or shape of the apparent
horizons, we use control theory to dynamically update the parameters of
the moving mesh periodically during the simulation. The time-dependent
coordinate maps of the moving mesh and control
signals used to update them are discussed in~\cite{Hemberger:2012jz} in
\S 4.1-4.3, \S 4.5, and \S 5 for the inspiral and \S 6 for the ringdown.

The details of how the control systems are implemented within the context of
asynchronous task-based parallelism along with the local adaptive time stepping
described in \S\ref{sec:time integration} are described in
\cite{Nelli:2025inprep}.

\section{Results\label{sec:results}}

In this section, we begin by testing \texttt{SpECTRE}'s long-term stability and
convergence; first with evolutions of single black holes in different coordinate
systems (\S\ref{sec:kerr black hole}) and then with an evolution of a
time-dependent gauge wave on a flat spacetime background (\S\ref{sec:gauge
wave}). Finally, we present results from a complete simulation of the inspiral,
merger, and ringdown of two black holes (\S\ref{sec:bbh results}). The
\texttt{SpECTRE} input files used for simulations, including generating the BBH
initial data, are provided as ancillary material with the paper.

\subsection{Single black hole evolutions\label{sec:kerr black hole}}

\begin{figure}
  \centering
  \includegraphics[width=0.3\linewidth]{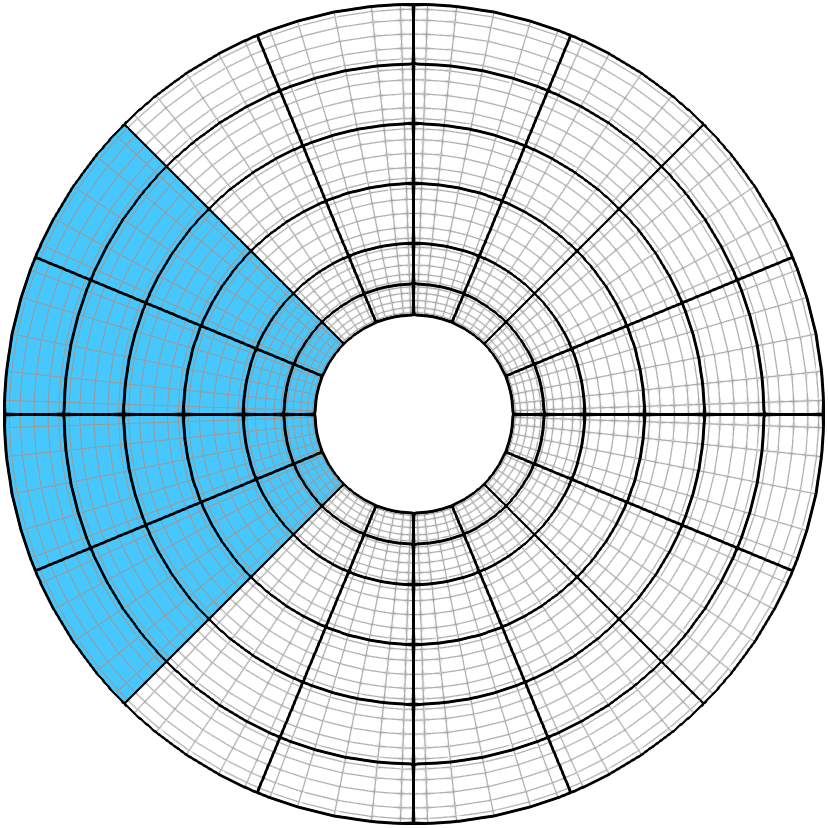}
\caption{An illustration of a slice through the computational domain used in the evolutions
of single black holes described in this paper. Six regions, each in the
shape of a deformed cube, combine to cover the volume of a spherical shell,
with the inner boundary (the excision surface) inside the black hole's 
apparent horizon. The six regions are themselves refined radially and 
angularly into smaller deformed cubes. One of the six regions is shown in blue
for clarity. \label{fig:bhdomain}}
\end{figure}

In this section, we use \texttt{SpECTRE} to evolve a single, stationary, black
hole that, unless otherwise noted, is non-spinning. We evolve a black hole from
analytic initial data corresponding to a black hole at rest centered at the
origin with zero spin. We choose the mass of the black hole to be $M=1$ and work
in units of $M$. In each evolution we use the following values for the FOGH
constraint damping parameters,
\begin{eqnarray}
  \gamma_0 &= \gamma_2 = A_0 e^{-r^2/w_0^2} + A_1 e^{-r^2/w_1^2},\\
  \gamma_1 &= - 1,
\end{eqnarray}
where $r$ is the coordinate distance from the origin, $A_0=7.0/M$, $A_1=0.1/M$,
$w_0 = 2.5 M$, and $w_1 = 100.0 M$. The computational domain of each evolution
covers a spherical shell volume (figure~\ref{fig:bhdomain}) with inner radius
$r_{\mathrm{in}}=r_{\mathrm{exc}}$ which differs for our different test cases,
and outer boundary coordinate radius
$r_{\rm out}=1,000M$. We apply boundary conditions as described
in \S~\ref{sec:boundary conditions}. We use a fourth-order Adams-Moulton
predictor-corrector time integrator with absolute and relative time stepper
tolerances of $10^{-8}$ and $10^{-6}$, respectively, unless otherwise stated.

\subsubsection{Kerr-Schild coordinates\label{sec:kerr-schild single hole}}

\begin{figure}[h]
  \includegraphics[width=0.49\columnwidth]{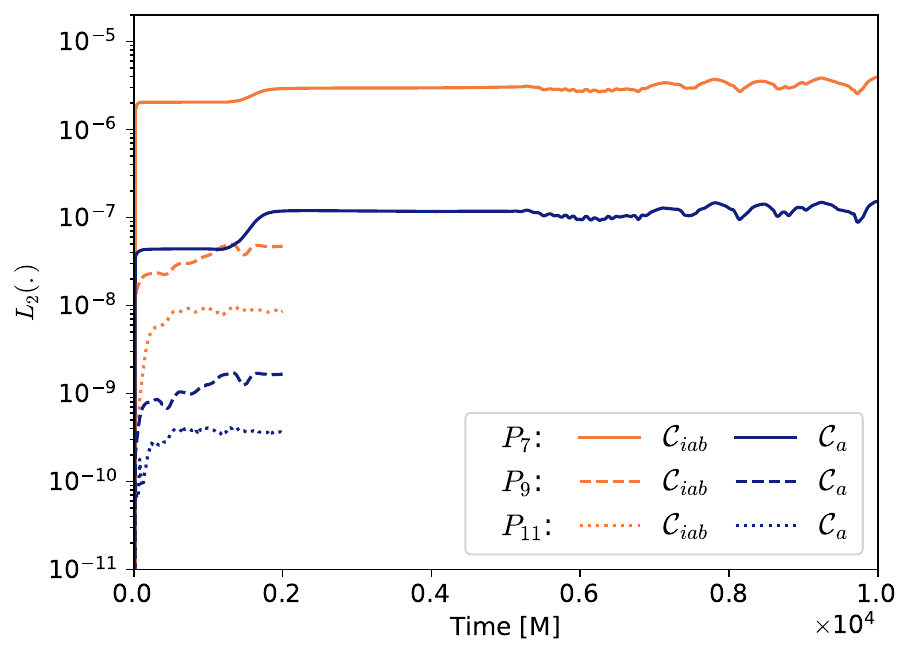}
  \includegraphics[width=0.49\columnwidth]{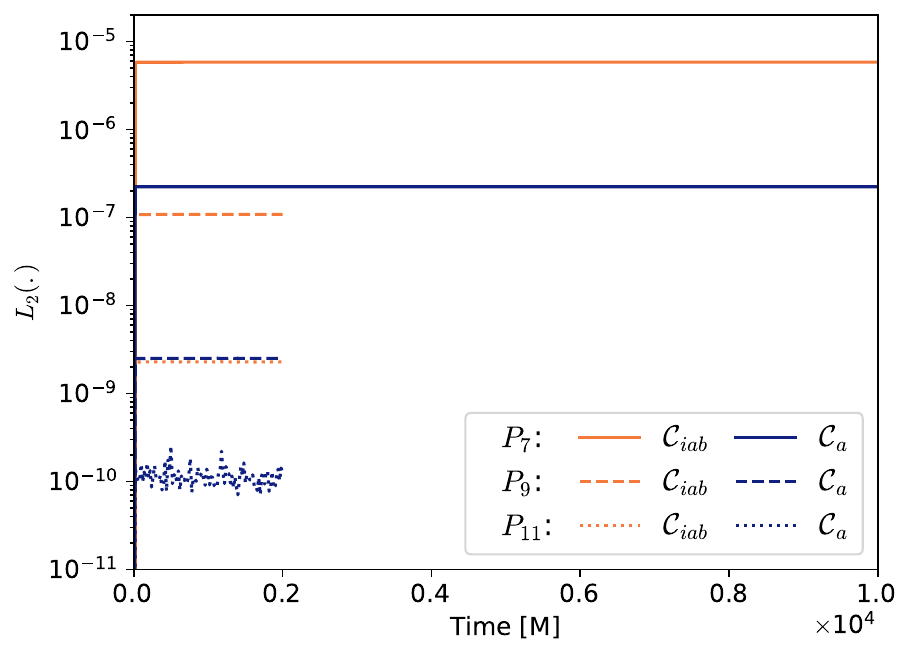}
  \caption{Constraint violations for single black hole evolutions in Kerr-Schild
    coordinates with Kerr-Schild initial data. We evolve the
    lowest resolution to time $t=10,000M$ to demonstrate long term stability and
    evolve two higher resolutions to time $t=2,000M$ to assess convergence with
    spatial resolution. \textit{Left}: Non-spinning black hole. \textit{Right}:
    Spinning black-hole with dimensionless spin $\chi=S/M^2 =0.8$.}
  \label{fig:bh-ks-conv}
\end{figure}

We first evolve a single black hole in Kerr-Schild coordinates from Kerr-Schild
initial data. The inner radius of the computational domain is
\response{$r_{\mathrm{in}}=r_{\mathrm{exc}}=1.9M$}. In this case there are no
coordinate dynamics, so a feedback control system is not necessary, though it is
enabled in the simulations presented here. The left panel of
figure~\ref{fig:bh-ks-conv} shows the gauge constraint $\mathcal{C}_a$ and the
3-index constraint $\mathcal{C}_{iab}$ as a function of time for several
different resolutions.  We evolve the lowest resolution to time $t=10,000M$ to
assess long-term stability, and we evolve the medium and high resolutions to
assess convergence. To limit the computational cost of these tests, we choose to
evolve the medium and high resolutions only to time $t=2,000M$. All simulations
are stable to time $t=2000 M$, and the lowest resolution remains stable to
$t=10,000M$.  The amount of violation of the gauge constraint $\mathcal{C}_a$
and the 3-index constraint $\mathcal{C}_{iab}$ is indicative of the overall
constraint violation in the simulation. In this evolution, the constraints
remain \response{approximately} constant, and they decrease exponentially with
increasing $p$-refinement (that is, increasing points per cell per dimension),
as expected. \response{We suspect the transient at $t=2000M$ results from
constraint violations reflecting off the outer boundary back to the interior.}

The right panel of figure~\ref{fig:bh-ks-conv} demonstrates long-term stability
and convergence for the same setup but with a black hole of dimensionless spin
$\chi\equiv S/M^2=0.8$ (with $r_{\mathrm{in}}=r_{\mathrm{exc}}=1.57M$).  Again,
we see that the constraints remain \response{approximately} constant and
converge exponentially with increasing $p$-refinement.

\subsubsection{Harmonic coordinates\label{sec:harmonic single hole}}

\begin{figure}[h]
  \includegraphics[width=0.49\columnwidth]{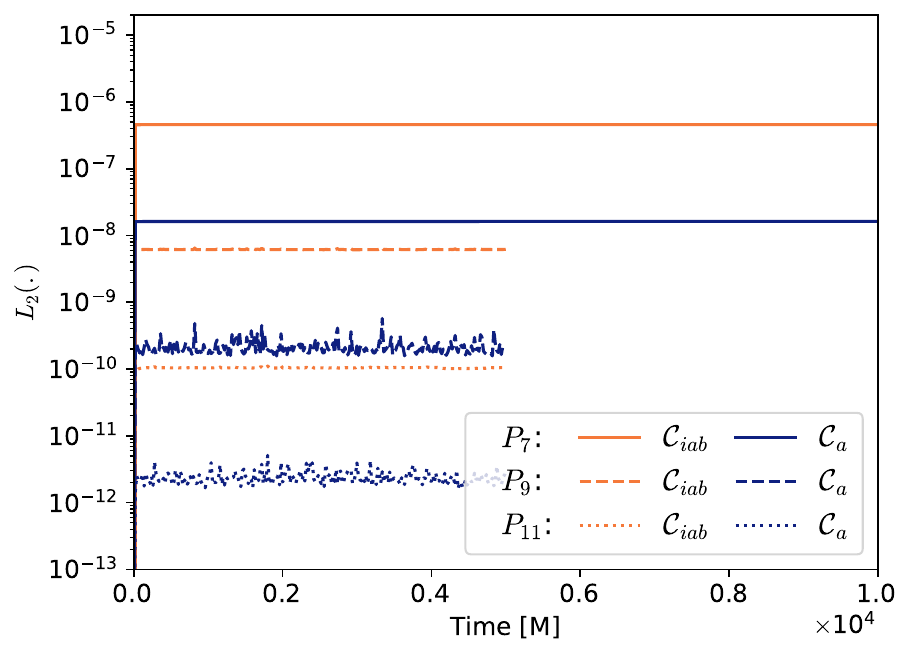}
  \includegraphics[width=0.49\columnwidth]{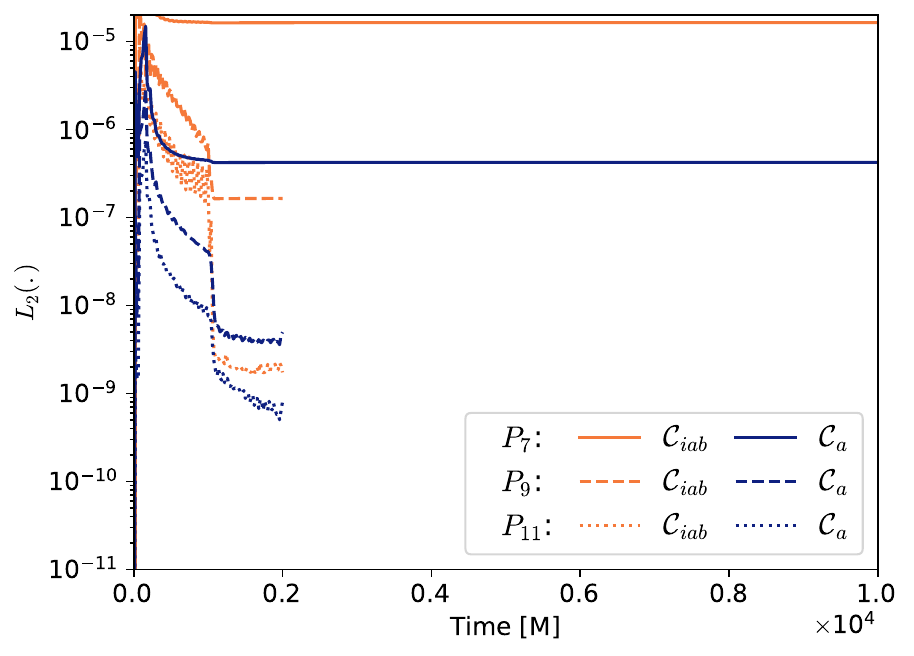}
  \caption{Constraint violations for single black hole evolutions in harmonic
    coordinates. We evolve the 
    lowest resolution to time $t=10,000M$ to demonstrate long term stability and
    evolve two higher resolutions to an earlier time to assess convergence with
    spatial resolution. \textit{Left}: Analytic initial data is in harmonic
    coordinates. Higher resolutions evolved to $t=5,000M$. The highest
    resolution has $10\times$ tighter time stepper tolerances.
    \textit{Right}: Analytic initial data is in Kerr-Schild coordinates. Higher
    resolutions evolved to $t=2,000M$. The difference between the initial data
    and evolution gauge causes non-trivial dynamics.}
    \label{fig:bh-harmonic-conv}
\end{figure}

Next, we evolve a single black hole in harmonic gauge $H_a=0$ using initial data
also in harmonic gauge. Here, $r_{\mathrm{in}}=r_{\mathrm{exc}}=0.9M$. The
absolute and relative time stepper tolerances for the highest resolution of this
test case are $10^{-10}$ and $10^{-8}$, respectively. Again,
since the initial data and evolution use the same gauge there are no gauge
dynamics. The left panel of figure~\ref{fig:bh-harmonic-conv} shows the gauge
constraint $\mathcal{C}_a$ and the 3-index constraint $\mathcal{C}_{iab}$ as a
function of time for several different resolutions. We evolve the lowest
resolution to time $t=10,000M$ to assess stability and two higher resolutions to
time $t=5,000M$ to assess convergence. The constraints again remain constant,
and they decrease exponentially with increasing $p$-refinement.

As a first test of the control system, we evolve a Kerr-Schild black hole in
harmonic coordinates. The inner radius of the domain again is
$r_{\mathrm{in}}=r_{\mathrm{exc}}=1.8M$. The differing gauge choices in the
initial data and evolution create coordinate dynamics that cause the BH horizon
to shrink. The control system (\S\ref{sec:control systems}) must decrease the
radius of the excision surface smoothly and precisely to avoid incoming
characteristic speeds, so that the problem remains well-posed and the code does
not terminate. The right panel of figure~\ref{fig:bh-harmonic-conv} shows
constraint violations over time for three resolutions. All evolutions are
stable, and the constraint violations converge away. However, the constraints
remain larger until after time $1,000M$. We suspect this is caused by initial
gauge dynamics, i.e., by time-dependent, outward-moving coordinate effects that
travel outward until exiting the domain through the outer boundary at
$r_{\mathrm{out}}=1,000M$.

\subsubsection{Damped harmonic coordinates\label{sec:dh single hole}}

\begin{figure}[h]
  \includegraphics[width=0.49\columnwidth]{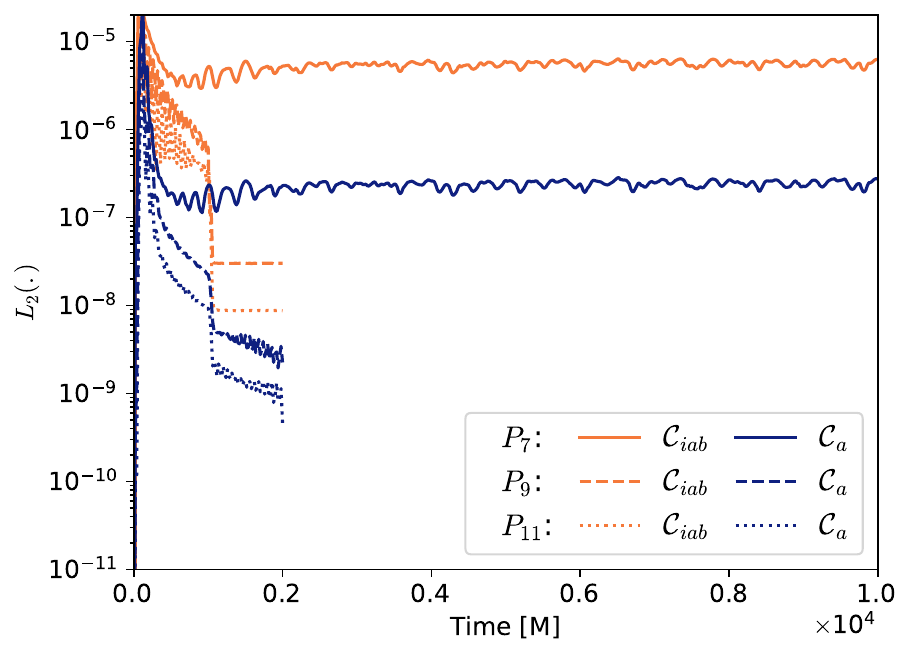}
  \includegraphics[width=0.49\columnwidth]{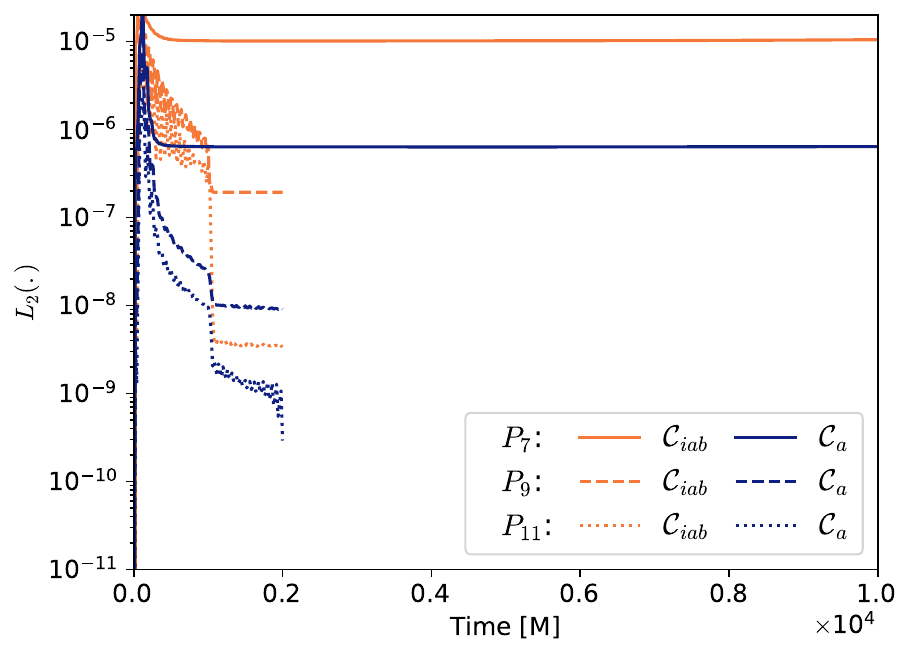}
  \caption{Constraint violations for single black hole evolutions in damped
    harmonic coordinates with Kerr-Schild initial data. We evolve the
    lowest resolution to time $t=10,000M$ to demonstrate long term stability and
    evolve two higher resolutions to time $t=2,000M$ to assess convergence with
    spatial resolution.  \textit{Left}: Non-spinning black hole. \textit{Right}:
    Spinning black-hole with dimensionless spin $\chi=S/M^2 =0.8$.}
  \label{fig:bh-dh-conv}
\end{figure}

Our final single-black-hole test consists of evolving Kerr-Schild analytic
initial data in damped harmonic gauge with
$r_{\mathrm{in}}=r_{\mathrm{exc}}=1.8M$.  The left panel of
figure~\ref{fig:bh-dh-conv} shows the constraints as a function of time for
several different resolutions. Just as in the harmonic gauge case, the
non-trivial gauge dynamics cause larger constraint violations until after one
light-crossing time to the outer boundary of $r_{\mathrm{out}}=1,000M$.  The
evolutions are 
stable and converge with increasing resolution. We also repeated this evolution
but for a black hole with a dimensionless spin of $\chi=0.8$ and
$r_{\mathrm{in}}=r_{\mathrm{exc}}=0.57M$.  We show the constraint violations in
the right panel of figure~\ref{fig:bh-dh-conv}. Again we see stable evolutions
and exponential convergence with increasing $p$-refinement.

\subsection{Gauge wave\label{sec:gauge wave}}
As a final test of convergence and stability, we evolve analytic initial data
consisting of a gauge wave in flat spacetime, a test conceived in
\cite{Alcubierre:2003pc} as part of a ``standard testbed'' for
NR codes. Physically, the solution is equivalent to flat
spacetime, but the chosen coordinates include a sinusoidal traveling wave that
introduces time-dependence, with a line element given by
\begin{equation}
  ds^2 = -H(t,x)\,dt^2 + H(t,x)\,dx^2 + dy^2 + dz^2,
\end{equation}
where
\begin{equation}
  H(t,x)=1 - A \sin\left(\frac{2\pi\left(x-t\right)}{d}\right),
\end{equation}
where $A$ and $d$ are the amplitude and wavelength of the gauge wave, which
travels along the $x$-axis.

We evolve analytic initial data of this solution, using the gauge source
function $H_a$ computed directly from the analytic initial data. We set the FOGH
constraint damping parameters to $\gamma_0=\gamma_2=1$ and $\gamma_1=-1$. We
evolve on the domain $[0, 1]^3$ with two elements in the $x$-direction, one
element in the $y$- and $z$-directions. We fix the $y$ and $z$ points per
element to 6 ($P_5$) and perform a convergence test by running three resolutions
with 15 ($P_{14}$), 18 ($P_{17}$), and 20 ($P_{19}$) points per element in the
$x$ direction. We apply periodic boundary conditions in all directions. We use a
sixth-order Adams-Moulton predictor-corrector time integrator. In our
simulations we choose $A=0.1$ and $d=1$.

\begin{figure}
  \includegraphics[width=0.49\columnwidth]{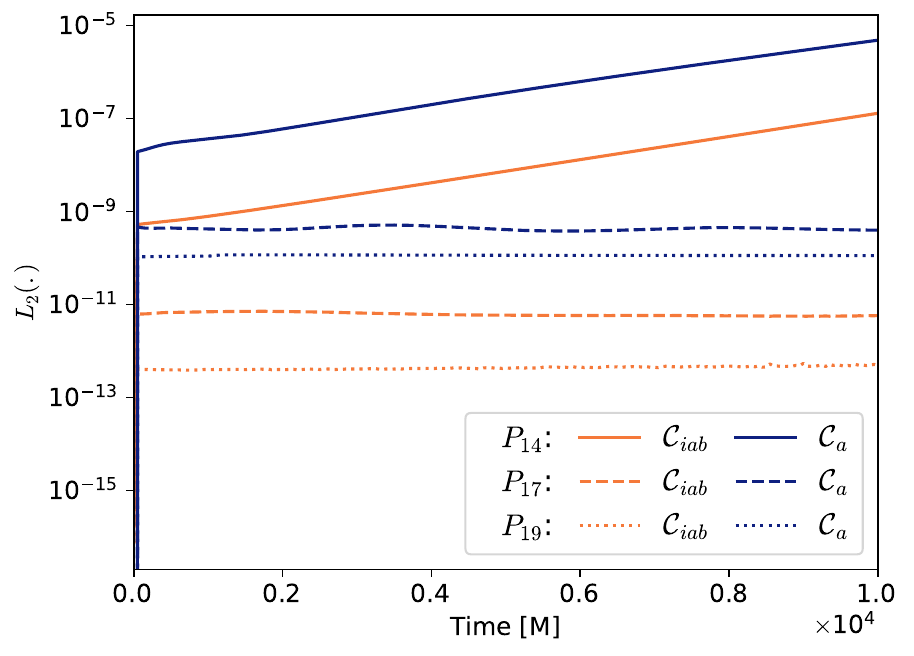}
  \includegraphics[width=0.49\columnwidth]{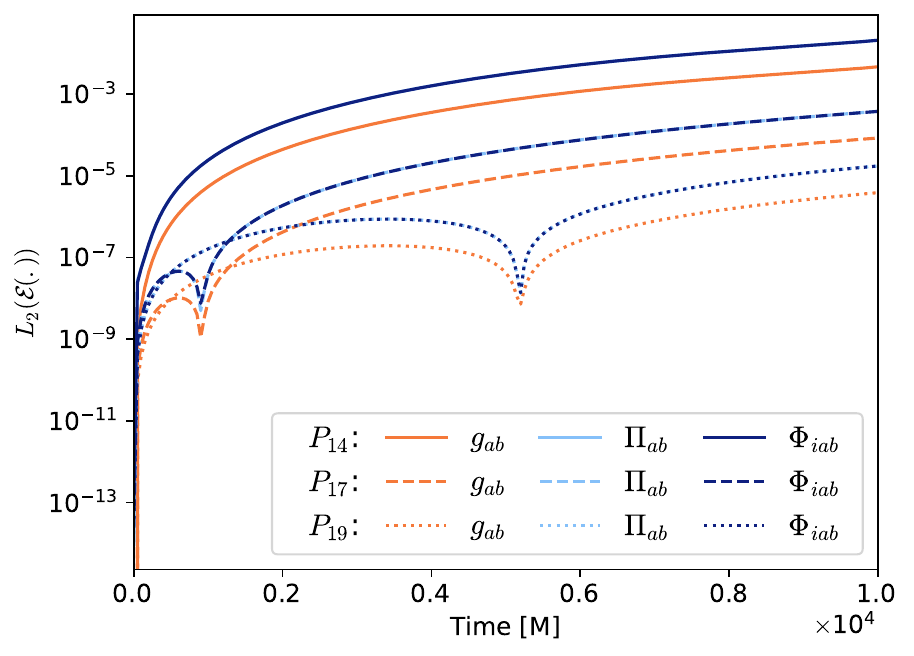}
  \caption{Results from an amplitude $0.1$ gauge wave simulation demonstrating
  long term stable and convergent simulations. We show simulations at three
  resolutions, 15, 18, and 20 grid points per element in the $x$-direction.
  {\it Left}: The 1-index and 3-index constraints as a function of time. The
  lowest resolution has exponentially growing constraints in time while the two
  higher resolutions have constant and convergent constraints in time.
  {\it Right}: \response{Difference (error) of the FOGH evolved variables
  $g_{ab}$, $\Pi_{ab}$, and $\Phi_{iab}$ from the analytic solution at the three
  resolutions. Since these are errors and not constraint violations, we expect
  the errors will accumulate linearly in time, even for constant constraint
  violations.} We see stable and convergent long-term behavior with the errors
  decreasing exponentially with increasing resolution. The errors for $\Pi_{ab}$
  and $\Phi_{iab}$ lie on top of each other and so only the errors for
  $\Phi_{iab}$ are visible.
  }\label{fig:nbh-gauge-wave-conv}
\end{figure}

Gauge wave simulations are known to be unstable in the BSSN formulation of the
Einstein equations~\cite{Babiuc:2007vr}, but are stable in the Z4
system~\cite{Alic:2011gg}. \response{Gauge wave simulations are stable in the FOGH system,
as we demonstrate with \texttt{SpECTRE} in figure~\ref{fig:nbh-gauge-wave-conv}}.
The left panel of figure~\ref{fig:nbh-gauge-wave-conv} shows the $L_2$ norm of
the 1-index and 3-index constraints as a function of time. While the lowest
resolution ($P_{14}$) simulation has exponentially growing constraints, the
higher resolution simulations have constant and convergent
constraints. Similarly, the right panel of figure~\ref{fig:nbh-gauge-wave-conv}
shows the $L_2$ norm of the error in the evolved variables $g_{ab}$, $\Pi_{ab}$,
and $\Phi_{iab}$ at the three resolutions. We observe stable and convergent
long-term behavior. The highest resolution simulation is close to being limited
by the time stepper tolerance.

\subsection{Binary black hole inspiral, merger, and ringdown\label{sec:binary
    black hole IMR}\label{sec:bbh results}}

In this section, we use \texttt{SpECTRE} to generate binary black hole initial
data and evolve the binary through $\sim18$ orbits of inspiral, merger, and
ringdown to a final, stationary state.  We then use \texttt{SpECTRE}'s Cauchy
Characteristic Evolution (CCE) module to evolve the outgoing gravitational waves
to future null infinity.  We perform the simulations at three different
resolutions, which we refer to as ``Lev0'', ``Lev1'', and ``Lev2'' with Lev0
being the lowest resolution and Lev2 being the highest. Each increase in
resolution increases the number of points per element per dimension by one.
During the inspiral, each simulation uses 4,800 elements, with Lev0, Lev1, and
Lev2 having $\sim2.6$; $\sim3.7$; and \response{$\sim5.0$} million total grid
points. During the ringdown, each simulation uses 7,680 elements and $\sim10.2$,
$\sim13.3$, and $\sim16.9$ million total grid points, respectively. All
simulations use the damped harmonic gauge to prevent the lapse collapsing and
$\sqrt{\gamma}$ diverging at merger. All simulations (both inspiral and
ringdown) also use a fourth-order Adams-Moulton predictor-corrector time
integrator with absolute and relative time stepper tolerances of $10^{-10}$ and
$10^{-8}$, respectively.

The evolutions were each performed on 10 compute-nodes in the Resnick High
Performance Computing Center at Caltech. Each compute node has two 28-core Intel
Cascade Lake CPUs. Our Lev0, Lev1, and Lev2 evolutions cost 58,000; 71,000; and
\response{117,000} core hours, which amounts to 104; 127; and \response{209}
wallclock hours, and an average of 120; 80; and \response{41} $M$/hour during
the inspiral.  In a future paper, we will assess \texttt{SpECTRE}'s performance
and scaling in more detail; our purpose for this paper is to demonstrate that
\texttt{SpECTRE} can evolve binary black holes through inspiral, merger, and
ringdown.

\subsubsection{Initial data\label{sec:bbh_id}} We begin our evolutions with
initial data of two equal-mass and non-spinning black holes in a quasicircular
orbit.  To generate the initial data we use the \texttt{SpECTRE} initial data
module~\cite{Fischer:2021voj,Vu:2021coj,Vu:2024cgf}, which solves the elliptic
constraint sector of the Einstein equations in the extended conformal thin
sandwich (XCTS) formalism~\cite{York:1998hy,Pfeiffer:2002iy,Pfeiffer2004-oo}. It
uses the superposed Kerr-Schild formalism to construct a conformal background to
the extended conformal thin sandwich equations based on the weighted
superposition of two isolated Kerr-Schild black
holes~\cite{Lovelace2008-sw,Lovelace:2010ne}. The black holes are represented as
excisions with negative-expansion apparent horizon boundary
conditions~\cite{Cook2004-yf,Varma2018-fp}. The initial data solver uses DG
methods similar to those described in this article to achieve scalable and
parallelizable solutions to the elliptic equations and is also open
source~\response{\cite{Vu:2021coj, spectrecode_2024}}.

\begin{figure}
  \centering
  \includegraphics[width=0.7\linewidth]{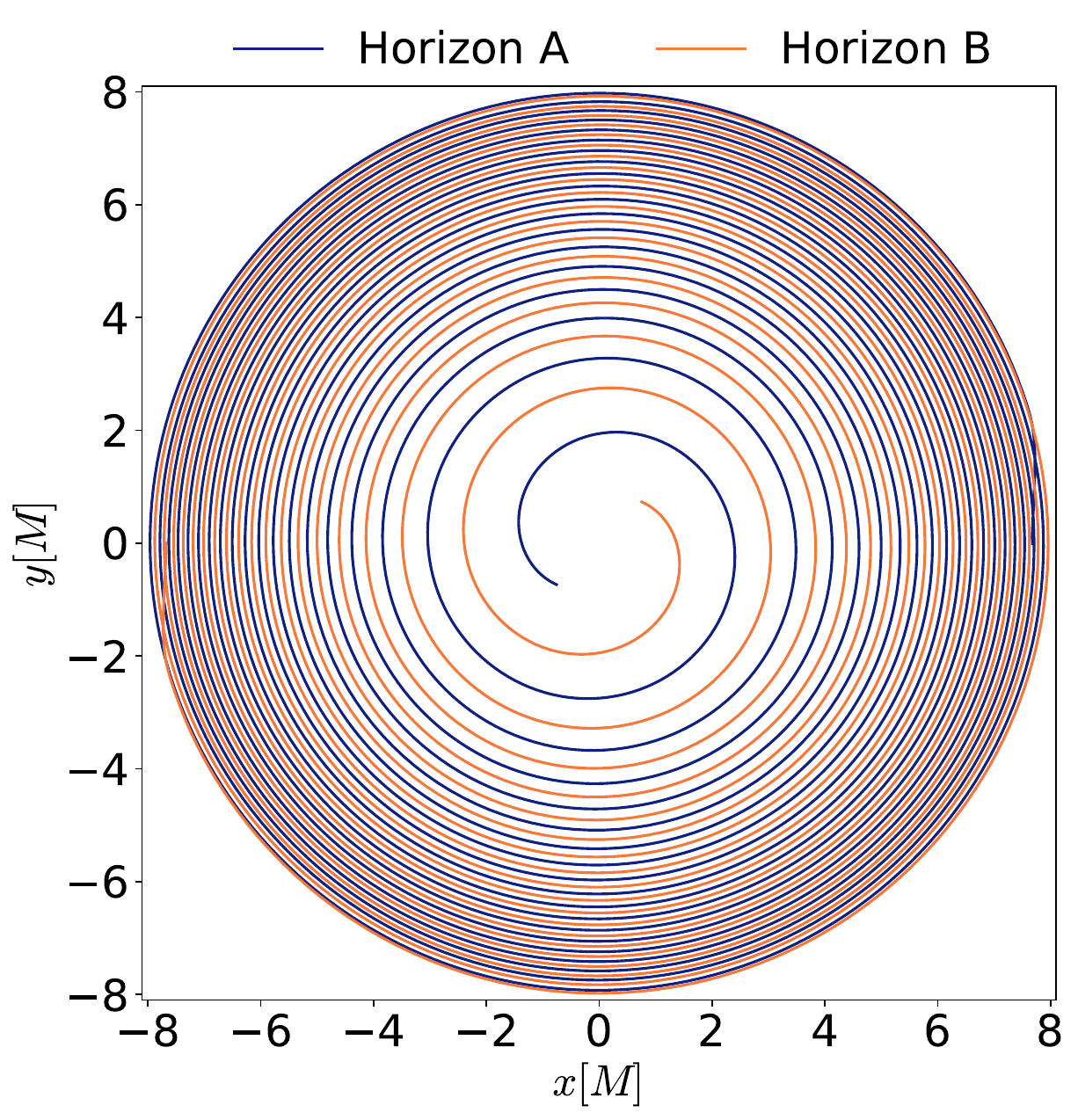}
  \caption{The trajectories of the centers of each apparent horizon during the
  Lev2 inspiral evolution, in the evolution's asymptotically inertial
  coordinates. The coordinates $x$ and $y$ indicate the position of the black
  hole within the orbital plane in units of the total initial mass $M$ of the
  binary.}\label{fig:trajectories}
\end{figure}

We choose an initial coordinate separation of the excision centers to be
$d_0=15.366M$ to facilitate future comparison with the family of simulations in
\cite{Hemberger:2013hsa} and to place the two black holes at $\sim18$ orbits
before they merge. The masses and spins of the black holes measured on the
horizons are driven to the desired values in a control loop that adjusts the
initial data parameters, similar to~\cite{Ossokine:2015yla}. In a second control
loop we performed eccentricity reduction by evolving the initial data for a few
orbits and adjusting the initial orbital parameters to iteratively reduce the
eccentricity of the orbit~\cite{Buonanno:2010, Habib:2024soh} to
\response{$7.2\times10^{-4}$}.  The resulting initial data parameters are
summarized in table~\ref{table:id}, and a plot of the resulting inertial
trajectories for Lev2 is shown in figure~\ref{fig:trajectories}.

\begin{table}
  \caption{Parameters used to generate the binary-black-hole initial data
    evolved in \S\ref{sec:bbh results}: in terms of the sum of the black holes'
    Christodoulou masses $M=M_1+M_2$ of each black hole, the individual masses
    $M_1$ and $M_2$, the dimensionless spin magnitudes $|\chi_1|/M^2$ and
    $|\chi_2|/M^2$, the orbital eccentricity $e$, the initial coordinate
    separation $d_0$ of the horizon centers, the initial orbital angular
    velocity $\Omega_0$, and the initial expansion rate $\dot{a}_0$.}
  \label{table:id}
  \begin{indented}
    \lineup
\item[]\hspace{-1.3em}\begin{tabular}{@{}cccccccc} \br
  $M_1/M$ & $M_2/M$ & $|\chi_1| / M^2$
  & $|\chi_2| / M^2$ & $d_0 / M$ & $M \Omega_0$ & $e$ &  $\dot{a}_0$ \\
  \hline
  0.5 & 0.5 & \response{$3 \times 10^{-8}$}
  & \response{$3 \times 10^{-8}$} & 15.366 & 0.0159 & \response{$7.2\times10^{-4}$}
                                                      & $3.4 \times 10^{-5}$ \\
    \br
  \end{tabular}
\end{indented}
\end{table}

\subsubsection{Computational domain decomposition}

\begin{figure}
  \raggedleft
  \begin{minipage}{0.43\columnwidth}
    \centering
    \includegraphics[width=\linewidth]{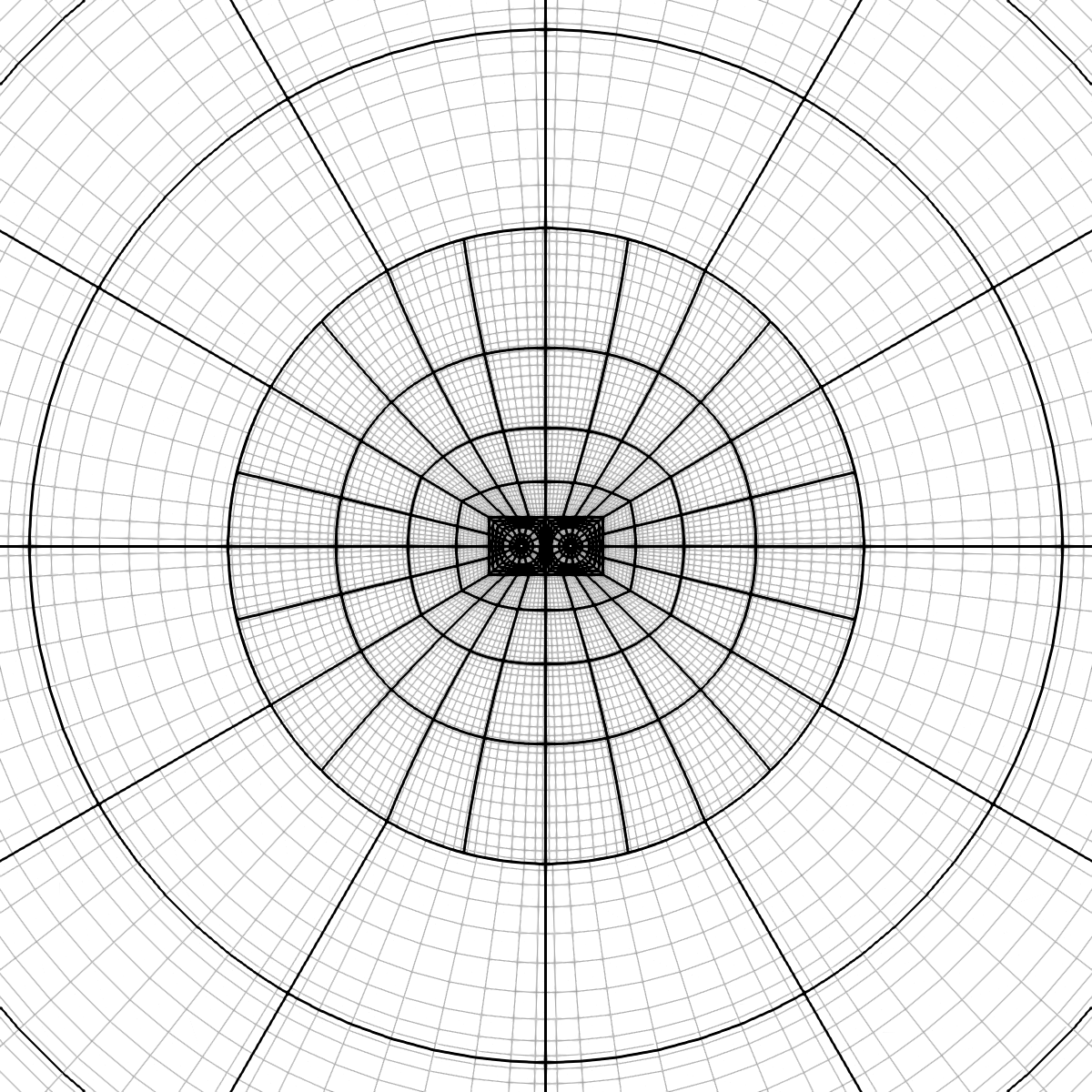}
    \\
  \end{minipage}
  \begin{minipage}{0.43\columnwidth}
    \centering
    \includegraphics[width=\linewidth]{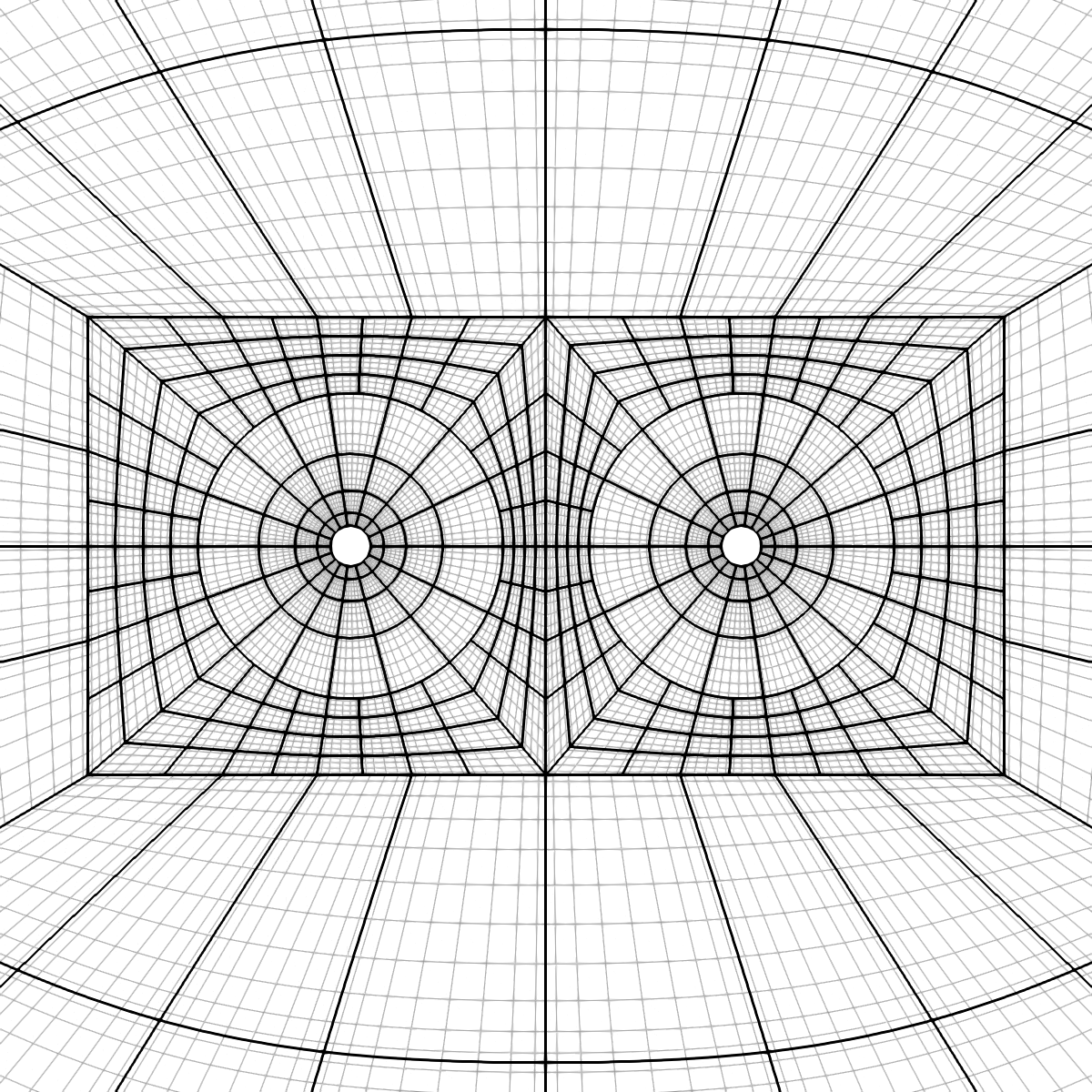}
    \\
  \end{minipage}
  \caption{An illustration of the computational grid used during the inspiral.
    We make use of two excision regions, 
    each region lying inside a black hole's apparent horizon. Each excision is
    surrounded by a spherical shell partitioned into six deformed cubes as in
    figure~\ref{fig:bhdomain}. Each spherical shell is then surrounded by another
    shell of six deformed cubes that transition to a cubical boundary. Then the
    two cubes themselves are surrounded by a transitionary envelope which becomes
    spherical. \textit{Left:} The transitionary envelope. \textit{Right:} A
    close-up of the domain structure around the excisions. The center of each
    excision is offset from the center of the cube.}\label{fig:bbhdomain}
\end{figure}

\begin{figure}
  \centering
  \includegraphics[width=0.4\linewidth]{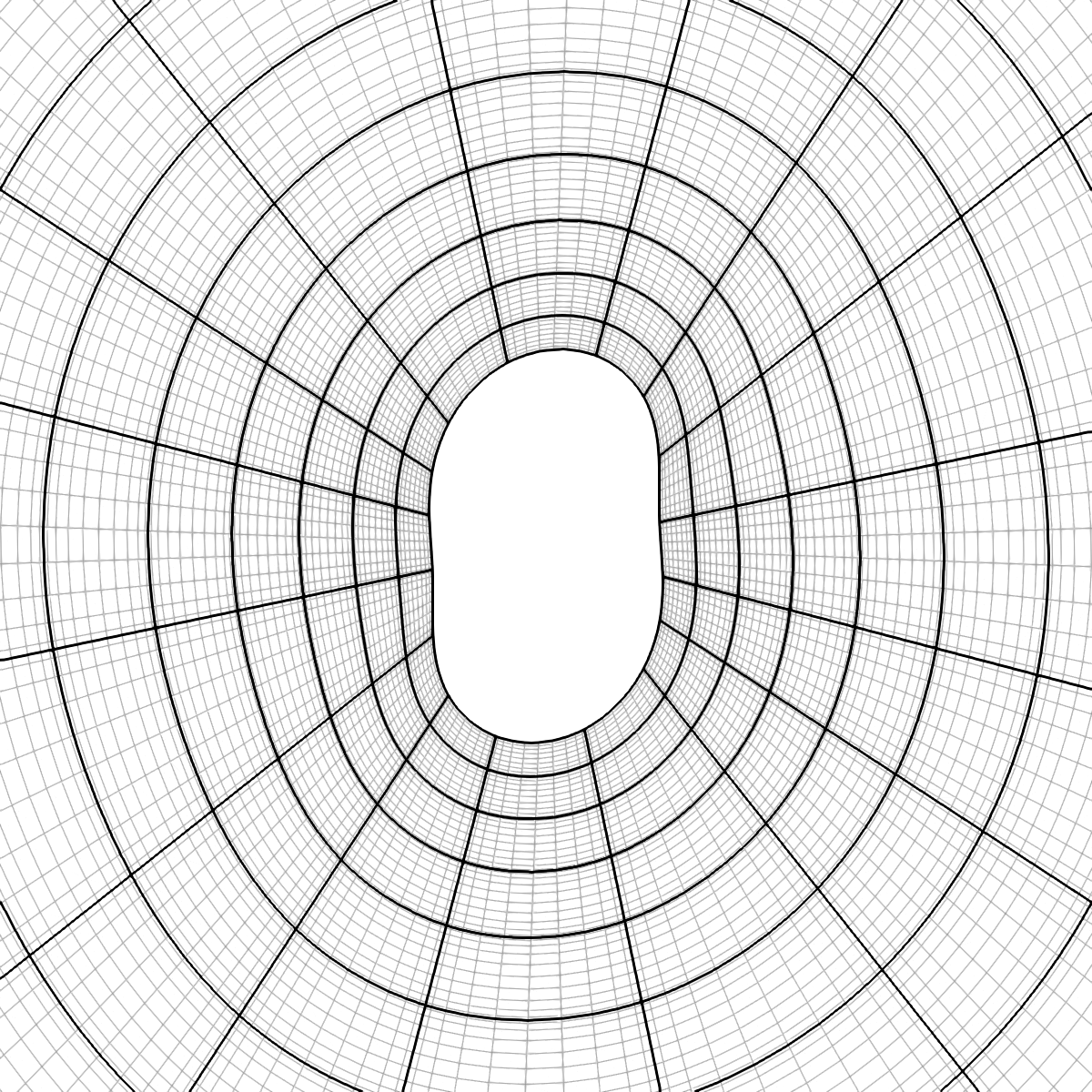}
  \caption{An image of the ringdown domain. After a common horizon is found,
    the evolution is regridded onto a domain with a single excision as in
    figure~\ref{fig:bhdomain}. Additional time-dependent maps are used to deform
    the excision's spherical shape into one that corresponds to the common
    horizon shape.
    \label{fig:ringdown_domain}
  }
\end{figure}

The numerical evolution of the FOGH system is performed with a DG scheme in which
the physical domain of the problem is partitioned into deformed hexahedral
elements with conforming boundaries. The boundaries and gridpoint distributions
of an element are determined by a continuous and differentiable coordinate map
applied to the logical Cartesian coordinates, which we label $\xi, \eta$ and
$\zeta$, of a regular cube spanning $[-1,1]^3$. The maps corresponding to
neighboring elements are required to be continuous but are not required to be
differentiable at
element boundaries. This provides the flexibility necessary to construct the
complicated domains needed for binary merger simulations using DG methods. An
example of the domain used during the inspiral is shown in
figure~\ref{fig:bbhdomain} and an example of the ringdown domain is shown in
figure~\ref{fig:ringdown_domain}. While the coordinate maps provide
significant flexibility, we found that instabilities arise if neighboring
elements differ by significantly more than a factor of two in size, placing a
practical constraint on how quickly the resolution can be reduced as one moves
away from the BHs.

Our computational domain is the region of space between the outer spherical
boundary and the excision boundaries that remove the black hole singularities
from the domain. The excision boundaries are spherical in the comoving
coordinates with their sizes and shapes in the inertial coordinates informed by
the size and shape of the apparent horizons (\S\ref{sec:control systems}). At
merger the apparent horizons of
the inspiraling black holes become enveloped by a single common apparent
horizon.  We handle the different number of excision boundaries during the
inspiral and ringdown by having distinct domains for each. At merger we
interpolate data from the inspiral domain that has two excision boundaries to
the ringdown domain that has one excision boundary.  We describe each domain
below.
\\

\noindent\textit{The inspiral domain:}

The inspiral domain is more complicated than the ringdown domain. This is
because of the complexity of having two excisions. During the
inspiral we must tile a two-excision domain with conforming hexahedra. Our
solution to this tiling problem makes use of 44 element collections grouped
into two radial and two ``biradial'' layers. Our description of the domain
decomposition starts at the excision surfaces and extends radially outward.

The first layer consists of six wedges composing the spherical shell surrounding
each excision. Each wedge is subdivided into multiple elements. The black holes
are located on the $x$-axis at $x\sim\pm7.683M$ with an excision radius of
$r_{\mathrm{exc}}\sim0.792M$. The shell around each black hole has an outer
radius of $6M$. For a fixed target accuracy, distributing the grid points
logarithmically in radius and equiangularly~\cite{Sadourny:1972, RONCHI199693}
in angles significantly reduces computational cost.

The second layer uses a set of wedges that wrap the shells around each black
hole in a cubical shell, as seen in figure~\ref{fig:bbhdomain}. A consequence
of the decreasing separation between the two black holes during the inspiral is
that the size of the excision within each cube grows as the simulation
progresses. Since we only deform the region inside the cubes to conform to the
shape of the apparent horizons, the simulation will fail if the apparent
horizons grow beyond the cube boundaries. We remedy this by decoupling the
excision center from the center of the cube (see the right panel of
figure~\ref{fig:bbhdomain}),
effectively increasing the size of the cube relative to the size of
the excision by a constant factor that is sufficient to keep the apparent
horizons within the cubes throughout the simulation. This generalized map is
crucial for robust inspiral and merger simulations.

The third layer consists of the elements surrounding the cubes around each black
hole (layers 1 and 2). We refer to this region, which extends from $r\sim18M$ to
$r=100M$, as the ``envelope''. This layer serves to transition the grid point
distributions from what is used near the black holes to the distribution that is
used in the wave zone. We use a logarithmic map in the radial direction and we
interpolate between a ``biradial'' equiangular map used for two excisions and a
``radial'' equiangular map suited for the spherical outer boundary.

The fourth and final layer is a spherical shell extending from the end of the
envelope ($r=100M$) to the outer boundary at $r=600M$. This shell uses a linear
distribution in the radial direction and an equiangular distribution in the
angular directions. Since the GW wavelength is constant in radius, a linear
distribution is necessary to avoid under resolving the waves. In production
quality simulations we expect to place the outer boundary at $r=1,500M$, since
in \response{typical} \texttt{SpEC} simulations we \response{find that this
radius is necessary to avoid a center-of-mass drift caused by the gauge boundary
condition}. Errors in the gauge boundary condition fall off as $\sim1/r^2$. We
find that the drift is larger for longer simulations, but this can be
compensated for with a larger outer boundary. Since we use CCE to extract the
gravitational waves, a large domain for wave extraction is not required,
\response{as it would be if we instead extrapolated the waves to spatial
infinity}.

In addition to the hexahedral maps that partition the domain, we also globally
apply rotation and expansion maps that track the angular, radial,
and center of mass motion of the binary system.

\response{
To characterize the errors in these coordinate mappings, we introduce the
diagnostic quantity (which we refer to as the ``jacobian diagnsotic'') as

\begin{equation}
  C_{\hat{i}} = 1 - \frac{\sum_i |\partial_{\hat{i}} x^i|}
    {\sum_i |D_{\hat{i}} x^i|},
\end{equation}
following the convention in \S\ref{sec:dg method} where hatted quantities are in
the logical coordinates and non-hatted quantities are in the inertial
coordinates. In the numerator, $\partial_{\hat{i}} x^i$ is the analytic Jacobian
provided by the analytic coordinate mappings (both time dependent and
time-independent) described in this section. In the denominator, $D_{\hat{i}}
x^i$ is the Jacobian computed by taking numerical derivatives of the inertial
coordinates in each logical direction. The sums are over all gridpoints. If
$C_{\hat{i}} = 0$, the analytic and numerical Jacobians are identical, meaning
that our coordinate mappings are perfectly represented by our
Legendre-Gauss-Lobatto DG scheme.
}
\\

\noindent\textit{The ringdown domain:}

The ringdown domain is a single excision domain used after a common apparent
horizon has formed. An example of this domain is shown in
figure~\ref{fig:ringdown_domain}. It is similar in structure to the domain used
in the single black hole tests in \S\ref{sec:kerr black hole}. We use a
logarithmic radial map from the excision surface to $r=50M$ and use a linear
spacing further away to resolve the gravitational waves. We used an equidistant
map instead of an equiangular map in the angular directions.

A significant challenge compared to the single black hole evolution is that the
time-dependent maps used during ringdown must be initialized from, and matched
to, the corresponding time-dependent maps in the inspiral. The rotation and
expansion maps are matched and then decay exponentially to being
time-independent. Most challenging are the shape and size maps. For the shape
map we perform a least squares fit in time to the spherical harmonic
coefficients of the common horizon \response{(and their time derivatives)} found during the
inspiral. We fit to 100 times, and then initialize the shape \response{of the excision} by
evaluating the fit at the transition time. For the size \response{of the excision}, we
manually specify an excision radius $r_{\mathrm{exc}} = 1.45$ and gave the
excision surface an initial outward \response{radial} velocity of 1.0; we adjusted these
choices by hand until the ringdown was able to begin successfully, with the
excision surface remaining inside the common apparent horizon while having all
characteristic characteristic speeds pointing out of the computational domain
(cf.~\ref{sec:boundary conditions}).

\response{For the simulations presented here, we carried out the transition from the
inspiral domain to the ringdown domain manually. First, when the coordinate
separation of the centers of the black holes' apparent horizons in the
asymptotically inertial frame fell to less than $2.38M$ (a value chosen by hand
to roughly correspond to the time that the common horizon first forms), we chose
to have the simulation begin outputting the evolution variables at every
gridpoint every $\Delta t = 0.01M$. Then, we chose to terminate the inspiral
portion of the simulation when the coordinate separation between the two black
holes decreased to less than $2.138M$. This value was chosen to be late enough
that it yielded enough distinct common horizon finds, but also early enough that
the constraint energy (cf. \S\ref{sec:constraint violations}) didn't grow too
large. In the future, we intend to implement the techniques that \texttt{SpEC}
uses to automate the transition process.}

\subsubsection{Constraint damping}

Based on our experience evolving binary black holes in \texttt{SpEC}, we use a
superposition of three Gaussians and a constant for $\gamma_0$ and $\gamma_2$,
and a single Gaussian plus a constant for $\gamma_1$. See~(\ref{eq:fosh gh
  metric conjugate evolution}) for how the constraint damping terms appear in the
evolution equations. The motivation for the different Gaussians is to increase
the constraint damping near each black hole, which requires the Gaussians to
move with the black holes as they inspiral. In \texttt{SpEC} and
\texttt{SpECTRE} we achieve this by making the Gaussians functions of the
comoving ``grid frame'' coordinates $x^{\bar{\imath}}$. As the black holes
inspiral, their coordinate radius increases in the comoving $x^{\bar{\imath}}$
coordinates, which means the width $w$ of the Gaussian must also increase by the
same amount. Increasing the width is achieved by dividing the width by the
expansion factor $E(t)$, which starts at $1$ and decreases as the black holes
inspiral. The specific form of the damping parameters we use is
\begin{eqnarray}
  \label{eq:gamma_0_2_gaussian}
  \gamma_0(t, x^{\bar{\imath}})
  &=& \gamma_2(t, x^{\bar{\imath}}) = C
      + \sum_{I=0}^2 A_I
      \exp\left[-\left(\frac{\bar{r}_I}{\bar{w}_I
      / E(t)}\right)^2\right],\\
  \label{eq:gamma_1_gaussian}
  \gamma_1(t, x^{\bar{\imath}})
  &=& C + A_{0} \exp\left[-\left(\frac{\bar{r}_0}{\bar{w}_{0}}\right)^2\right],
\end{eqnarray}
where $C$ is a constant, $A_I$ are the amplitudes of the Gaussians, $\bar{r}_I$
are the grid-frame radii from the center of each Gaussian, and $\bar{w}_I$ are
the widths of the Gaussians in the grid
frame. Table~\ref{table:inspiral_damping} shows the parameters during the
inspiral and table~\ref{table:ringdown_damping} shows them during the
ringdown. In the grid frame the black holes are always located on the
$\bar{x}$-axis so we only specify the grid frame $\bar{x}$-coordinate at which
each Gaussian is centered, denoted by $\bar{x}^C$ in
tables~\ref{table:inspiral_damping} and~\ref{table:ringdown_damping}.

\begin{table}
  \caption{\label{table:inspiral_damping} Parameters for the Gaussians that make
    up the constraint damping functions during the
    inspiral. See~(\ref{eq:gamma_0_2_gaussian}) and~(\ref{eq:gamma_1_gaussian})
    for how the coefficients appear in the functional form of the constraint
    damping parameters.}
  \begin{indented}
    \lineup
  \item[]\hspace{-0.0em}\begin{tabular}{@{}ccccccccccc} \br & $C$ & $A_0$ &
      $\bar{w}_0$ & $\bar{x}_0^C$ & $A_1$ & $\bar{w}_1$ & $\bar{x}_1^C$
    & $A_2$ & $\bar{w}_2$ & $\bar{x}_2^C$ \\
    \hline $\gamma_0$ & $0.01/M$ & $0.75/M$ & 38.415 & 0 & $8/M$ & 3.5 & 7.683
    & $8/M$ & 3.5 & -7.683 \\
    $\gamma_1$ & $-0.999/M$ & $0.999/M$ & $10 d_0$ & -- & -- & -- & --
    & -- & -- & -- \\
    $\gamma_2$ & $0.01/M$ & $0.75/M$ & 38.415 & 0 & $8/M$ & 3.5 & 7.683
    & $8/M$ & 3.5 & -7.683 \\
    \br
  \end{tabular}
\end{indented}
\end{table}

\begin{table}
  \caption{\label{table:ringdown_damping} Parameters for the Gaussians that make
    up the constraint damping functions during the
    ringdown. See~(\ref{eq:gamma_0_2_gaussian}) and~(\ref{eq:gamma_1_gaussian})
    for how the coefficients appear in the functional form of the constraint
    damping parameters.}
  \begin{indented}
    \lineup
  \item[]\hspace{-0.0em}\begin{tabular}{@{}ccccccccccc} \br
  & $C$ & $A_0$ & $\bar{w}_0$ & $\bar{x}_0^C$
  & $A_1$ & $\bar{w}_1$ & $\bar{x}_1^C$
  & $A_2$ & $\bar{w}_2$ & $\bar{x}_2^C$ \\
  \hline
  $\gamma_0$ & $0.01/M$ & $1.0/M$ & 100 & 0
  & $7/M$ & 2.5 & 0
  & -- & -- & -- \\
  $\gamma_1$ & $-0.999999/M$ & -- & -- & --
  & -- & -- & --
  & -- & -- & -- \\
  $\gamma_2$ & $0.001/M$ & $0.1/M$ & 100 & 0
  & $7/M$ & 2.5 & 0
  & -- & -- & -- \\
  \br
  \end{tabular}
\end{indented}
\end{table}

\subsubsection{Constraint violations\label{sec:constraint violations}}

\begin{figure}
  \centering
  \includegraphics[width=0.8\linewidth]{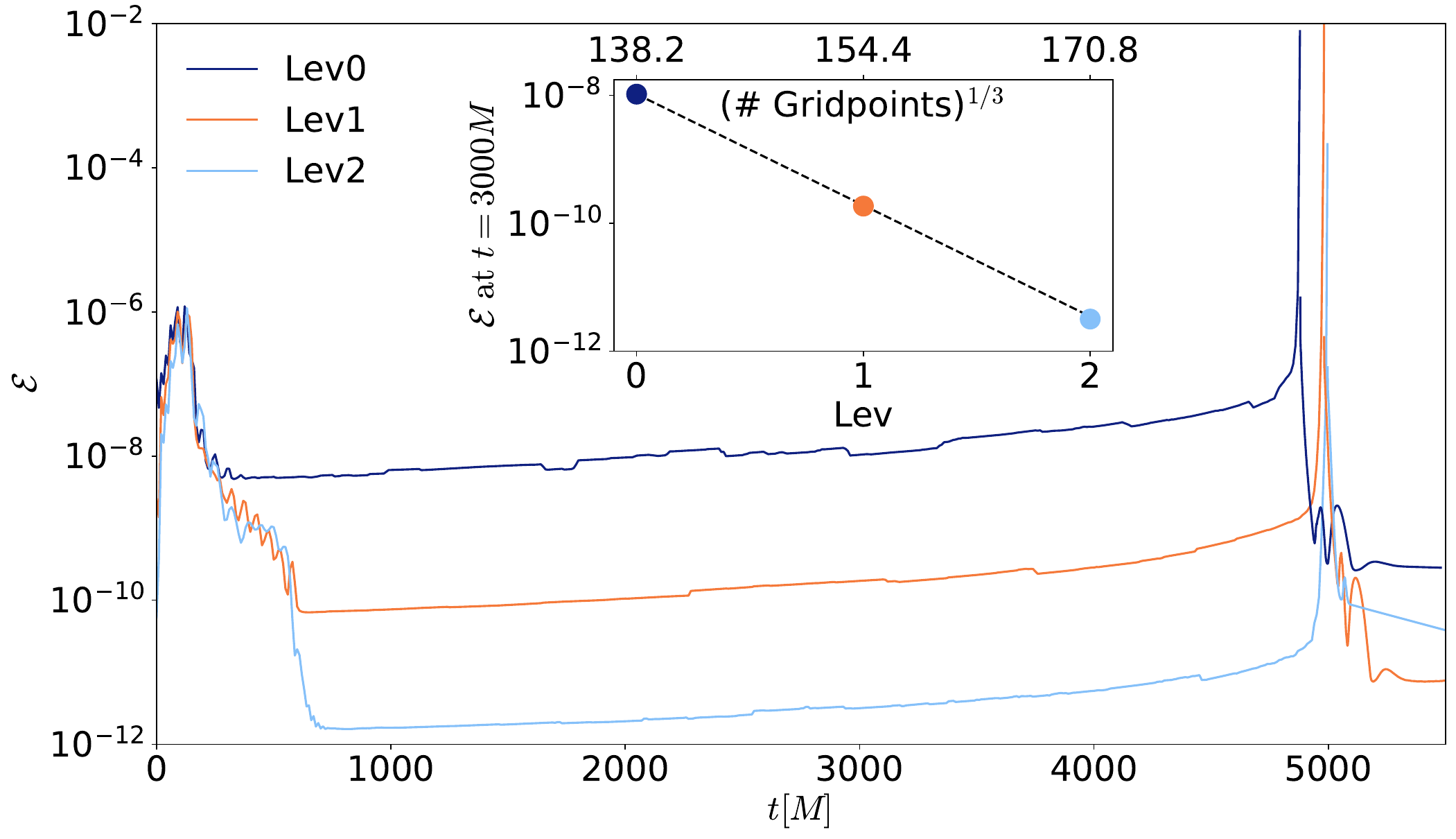}
  \caption{L2-norm of the constraint energy $\mathcal{E}$
  (see~(\ref{eq:constraint energy})) over the entire computational domain for
  evolutions of an equal-mass, non-spinning, binary black hole at low
  (``Lev0''), medium (``Lev1''), and high (``Lev2'') spatial resolution, as a
  function of time $t$ in units of the initial total mass $M$. The peak of the
  constraint energy for each resolution is when we transition to the ringdown
  grid. We see exponential convergence with resolution except at merger and
  early ringdown, where \response{we suspect} adaptive mesh refinement will be
  necessary to resolve the additional dynamics. \response{The inset shows this
  exponential convergence at time $t=3000M$ during the inspiral as a function of
  the cube root of the total number of grid points (and ``Lev'').}
  \label{fig:bbh-constraints}}
\end{figure}

Figure~\ref{fig:bbh-constraints} shows the constraint energy $\mathcal{E}$
(see~(\ref{eq:constraint energy})) as a function
of time at each resolution. Experience from \texttt{SpEC} suggests that the
initial, rapid growth in $\mathcal{E}$ is caused by not resolving the initial
data's rapid relaxation and emission of spurious, high-frequency ``junk''
gravitational radiation at the start of the simulation. Resolving the junk
radiation is computationally expensive and generally not done in NR evolutions
of binary black holes. After the initial growth of constraint violation damps
away, the constraints converge exponentially with increasing $p$-refinement.
The constraints grow sharply near the time of merger, as the black holes become
more distorted by each others' tidal gravity, causing the solution to be less
resolved by our fixed computational mesh. We anticipate that future
\texttt{SpECTRE} simulations using adaptive mesh refinement will improve the
behavior of the constraints near the time of merger.

\response{
To ensure the errors from our coordinate mappings do not affect our constraint
violations, we compare the L2-norms of the constraint energy and the jacobian
diagnostic in table \ref{table:mapping_error} at $t=3000M$ during the inspiral.
For all our resolutions, the mapping error is below our constraint violations,
meaning that our coordinate maps do not affect the results of these simulations.
}

\begin{table}
  \caption{\label{table:mapping_error} \response{The L2-norm of the constraint energy
  $\mathcal{E}$ and jacobian diagnostic $C_{\hat{i}}$ over all gridpoints at
  $t=3000M$ during the inspiral. For all of our resolutions, the jacobian
  diagnostic (which is a measure of our mapping error) is below the error of the
  constraint energy.}}
  \begin{indented}
    \lineup
  \item[]\hspace{-0.0em}\begin{tabular}{@{}cccc} \br
  L2-norm & Lev0    & Lev1    & Lev2\\
  \hline
  $\mathcal{E}$ & $1.0 \times 10^{-8}$  & $1.9 \times 10^{-10}$ &
  \response{$3.2 \times 10^{-12}$}\\
  $C_{\hat{i}}$ & $8.8 \times 10^{-11}$ & $6.2 \times 10^{-12}$ &
  \response{$5.7 \times 10^{-13}$}\\
  \br
  \end{tabular}
\end{indented}
\end{table}

After the merger, as the remnant black hole rings down, the constraint
violations decrease to much smaller values again. When the black hole has
settled to its final stationary state, the constraint violations continue to
slowly decrease in time. Because the ringdown constraints in the highest
resolution are not smaller than those with the medium spatial resolution, we
suspect that the numerical error is dominated not by spatial resolution in the
ringdown but by some other factor. One possibility is the time-stepping accuracy
during the late inspiral and ringdown. We leave a careful study of this,
including improvements to the domain decomposition and use of adaptive mesh
refinement during the ringdown, to future work.

\subsubsection{Apparent horizons}

Another way of measuring the accuracy of BBH simulations is to track the masses
and spins of the BHs using the apparent horizon surfaces. \response{We find the apparent
horizons surfaces using a ``fast flow'' approach similar to the one outlined in
\cite{Gundlach:1997us}. There are a number of subtleties that arise when
implementing this approach with an asynchronous task-based parallelism setting.
These are due to the fact that the ``fast flow'' approach is an iterative method
and thus needs to store the metric on the full computational domain at a given
simulation time until all iterations are complete. A more in-depth explanation
of this will be available in \cite{Nelli:2025inprep}.

We measure masses and spins on the individual apparent
horizons every $0.5M$ during the inspiral and merger and on the common apparent
horizon every $0.1M$  during ringdown.} The irreducible mass, defined as
$M_{\mathrm{irr}}\equiv \sqrt{A/16\pi}$, where $A$ is the surface area of the
apparent horizon, should be monotonically increasing. Thus, any decreases in
$M_{\mathrm{irr}}$ can be viewed as a measure of the numerical error in the
simulation. Another useful metric is the Christodoulou mass
$M_{\mathrm{Ch}}\equiv \sqrt{M_{\mathrm{irr}}^2 + S^2/4M_{\mathrm{irr}}^2}$,
which includes both the irreducible mass and rotational kinetic energy. The
dimensionless spin $\chi = S/M_{\mathrm{Ch}}^2$ measures the spin in terms of
approximate rotational Killing vectors, as discussed in Appendix A of
\cite{Lovelace2008-sw}. For equal-mass non-spinning simulations
$M_{\mathrm{irr}}$ and $M_{\mathrm{Ch}}$ should remain constant until merger,
while $\chi$ should remain identically zero. Deviations from this behavior help
quantify numerical errors in the simulation.

\begin{figure}
  \hfill
  \includegraphics[width=0.9\linewidth]{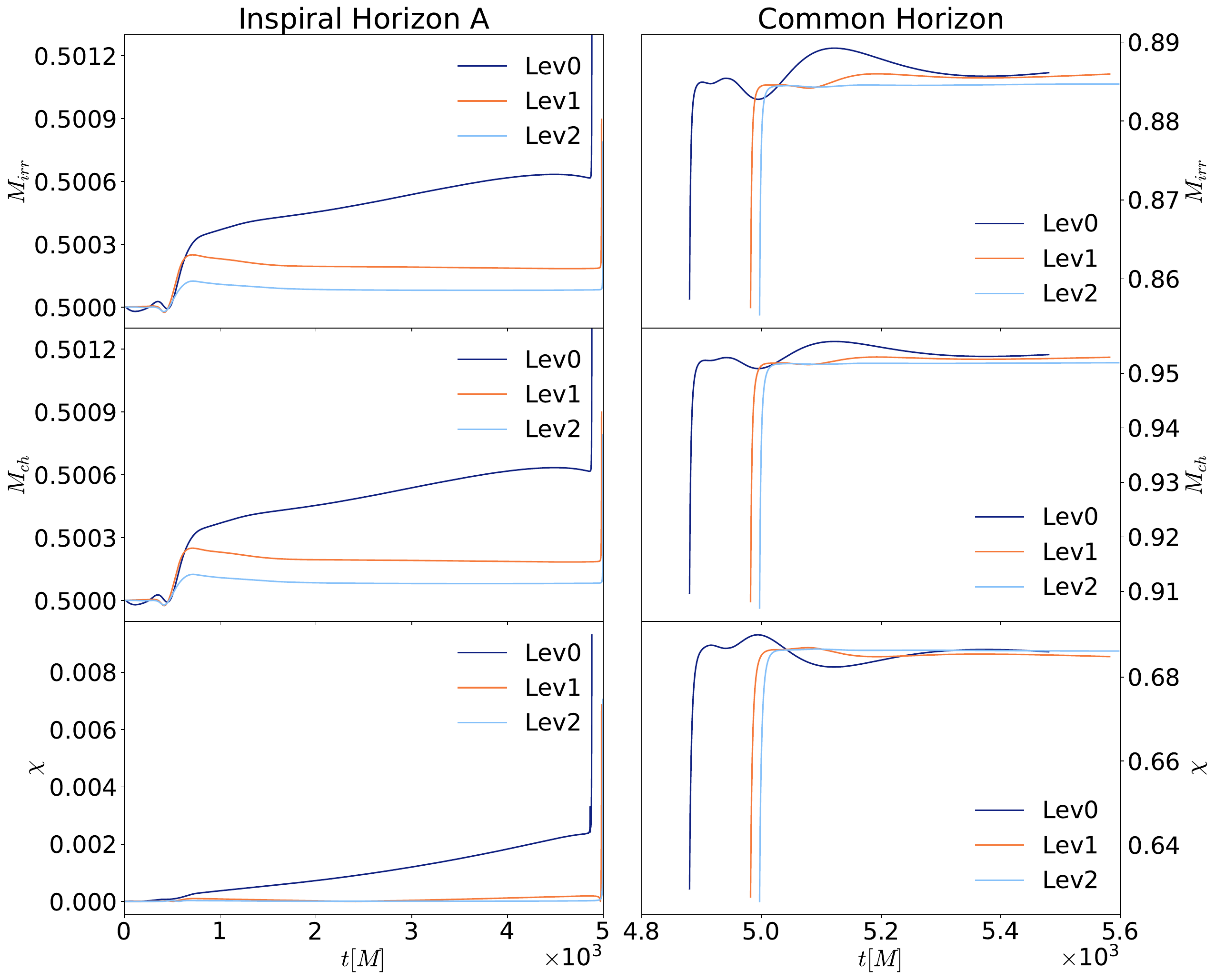}
  \caption{Apparent horizon masses and spins as a function of time. The left
    column shows the irreducible mass $M_{\rm irr}$, Christodoulou mass
    $M_{\rm ch}$, and dimensionless spin angular momentum
    $\chi\equiv S/M_{\rm ch}^2$ for one of the individual apparent horizons
    during inspiral, as
    a function of time $t$ in units of the total initial Christodoulou mass
    $M$. The right column shows the same quantities during the ringdown for the
    common apparent horizon. We see convergence with increasing resolution,
    specifically that the Lev1 and Lev2 evolutions track each other more closely
    than the Lev0 and Lev1 simulations.\label{fig:massesAndSpins}}
\end{figure}

Figure~\ref{fig:massesAndSpins} shows $M_{\mathrm{irr}}$, $M_{\mathrm{Ch}}$, and
$\chi$ during the inspiral and ringdown. We find that the masses and spins over
time are convergent, in the sense that the difference between Lev0 and Lev1 is
greater than the difference between Lev1 and Lev2. During the inspiral, after
the initial transient, the masses and spins remain more constant in time as
resolution increases, until sharp gains near the time of merger as the black
holes gain energy and angular momentum. During the ringdown, the masses and
spins relax to final, constant values, as expected. The final Christodoulou mass
$M_{\rm ch}=0.952$ differs from the initial Christodoulou mass ($M_{\rm ch}=1$)
by 4.8\%, and the final spin is $0.686$. Both values are consistent with the
fitting formulas in~\cite{Hemberger:2013hsa}, tuned using \texttt{SpEC}
evolutions of equal-mass, equal-aligned-spin binary black holes.

\subsubsection{Gravitational waveforms}
We compute gravitational waveforms using Cauchy-Characteristic Evolution
(CCE)~\cite{1996PhRvD..54.6153B,2016LRR....19....2B, 2016CQGra..33v5007H,
2020PhRvD.102b4004B, 2020PhRvD.102d4052M}, using the \texttt{SpECTRE}
implementation of CCE~\cite{Moxon:2021gbv}. This method utilizes an additional
characteristic evolution code, the one described in~\cite{Moxon:2021gbv}, that
solves the full Einstein equations on a set of outgoing null slices that extend
from some inner worldtube all the way to future null infinity.  Boundary
conditions on the worldtube are provided by the interior Cauchy evolution, in
this case also done with \texttt{SpECTRE}.  For the characteristic evolution,
there is freedom to choose one complex function on the initial null slice, which
encodes the initial incoming radiation. We set it according to equation (16) of
\cite{Moxon:2021gbv}. From the characteristic evolution, one can compute the
gravitational-wave strain and all five Weyl scalars at future null infinity.
Gravitational waveforms computed via CCE are in a well-defined gauge modulo
Bondi-van der Burg-Metzner-Sachs (BMS)
transformations~\cite{Bondi:1962px,Sachs:1962wk}, which are extensions of
Poincar\'e transformations and correspond to symmetries of asymptotically flat
spacetimes at future null infinity. The raw output from CCE is in an effectively
random BMS frame, so to completely fix the gauge, it is necessary to perform a
BMS transformation~\cite{Mitman:2024uss}.  We choose to transform waveforms into
the superrest frame of the inspiral~\cite{Mitman:2021xkq,Mitman:2022kwt}, which
can be thought of as the BMS extension of a frame in which the binary is at rest
during the inspiral. We find the BMS transformation to map to this frame using
data from the strain and Weyl scalars over the time window $[1800M,2200M]$. See
\cite{Mitman:2024uss} and references therein for an in-depth review of BMS frame
fixing.

\begin{figure}
  \hfill
  \includegraphics[width=0.9\linewidth]{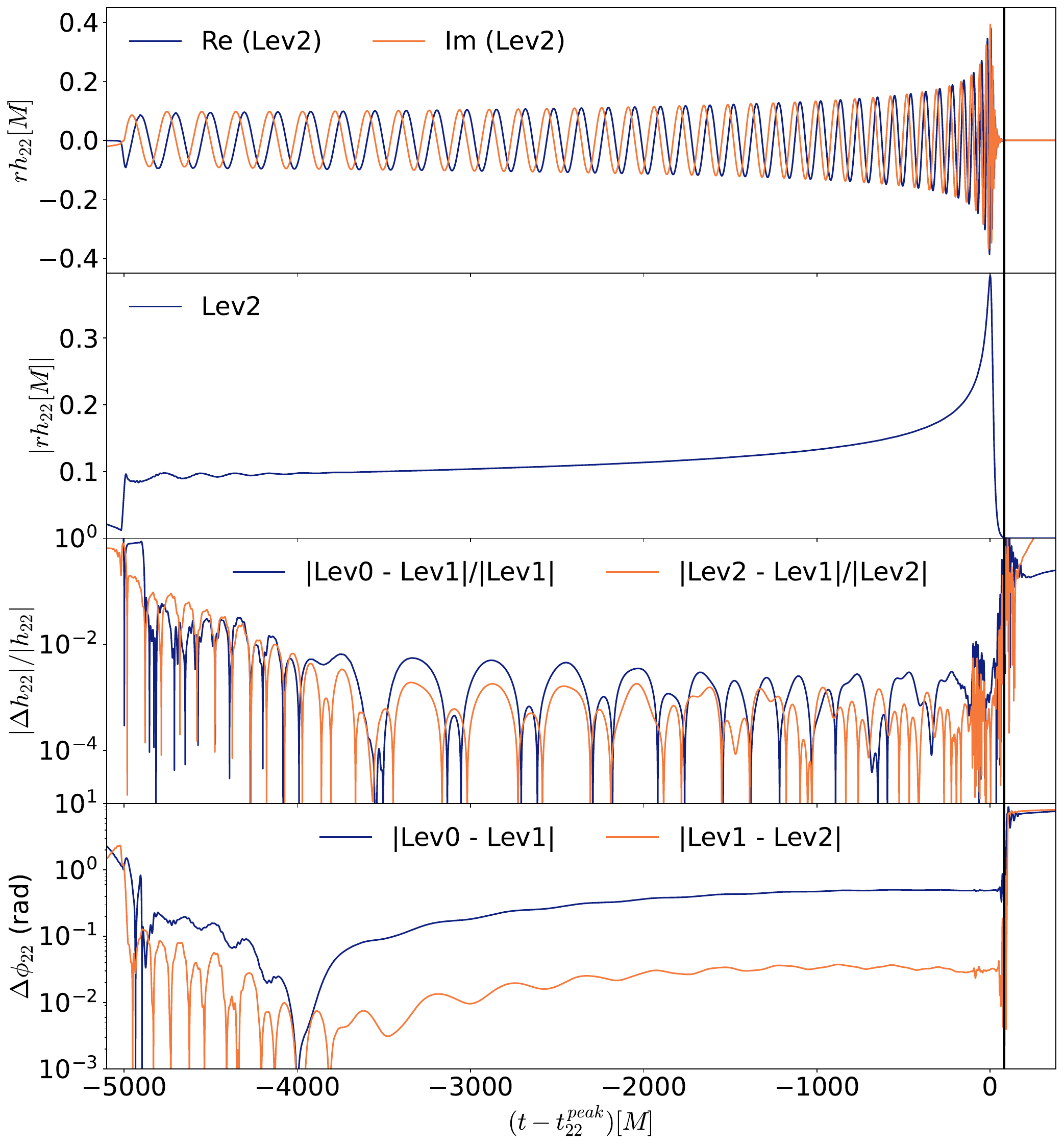}
  \caption{The $\ell=m=2$ spin-weighted spherical-harmonic mode of the
    gravitational waves, computed using Cauchy Characteristic Evolution (CCE)
    and frame-fixed using the Python package \texttt{scri}, with a worldtube
    radius of $200M$, where $M$ is the total initial Christodoulou mass. From
    top to bottom, as a function of retarded time $t-t^{\rm peak}_{22}$, the
    panels show i) the real and imaginary part of the strain $rh_{22}/M$; ii)
    the amplitude $|r h_{22}/M|$ at the highest spatial resolution; iii) the
    fractional amplitude difference $|\Delta h_{22}|/|h_{22}|$, defined as the
    magnitude of the $\ell=m=2$ amplitude difference between two spatial
    resolutions, divided by the magnitude of the higher spatial resolution's
    amplitude; iv) the difference in phase $\Delta \phi_{22}$, defined as the
    difference between the $\ell=m=2$ gravitational-wave phase at two different
    spatial resolutions, with each phase offset such that the phase vanishes at
    $t - t_{22}^{\rm peak} = -4000 M$. The vertical black lines shows where the
    amplitude goes below $\sim10^{-3}$. \label{fig:BbhGWs}}
\end{figure}

Figure~\ref{fig:BbhGWs} shows the leading-order $\ell=m=2$ spin-weighted
spherical harmonic mode of the gravitational-wave strain as a function of
retarded time $t$. To estimate the accuracy of the waveforms, we first apply a
time shift $t_{22}^{\rm peak}$ so that the peak amplitudes at each spatial
resolution occur at the same shifted time. Then, we apply a constant phase
offset such that the gravitational-wave phase of the $\ell=m=2$ mode vanishes at
time $t-t_{22}^{\rm peak}=-4000M$. This time was chosen as an early time after
most of the initial spurious ``junk'' radiation (especially visible in the
amplitude at early times) has been emitted. This junk radiation is
characteristically different than the junk radiation seen in
figure~\ref{fig:bbh-constraints}. The junk radiation in figure~\ref{fig:BbhGWs}
is of a lower frequency and is due to our choice of data on the initial null
slice. It is an active area of research to improve data on the initial null
slice.

We choose to post-process the data in this way because it enables us to
transform each simulation to some reasonable BMS frame without using information
from the other simulations. As a result, we can perform more meaningful
convergence tests, since the output from each simulation is independent of every
other. While one could obtain better agreement between different resolutions by
finding the BMS transformation which minimizes the residual between the
simulations' waveforms, this would go beyond computing a convergence error,
because the frame of one simulation is determined by the other.

Finally, we interpolate the amplitudes and phases at each spatial resolution
onto a common set of times and take differences, estimating the numerical error
of a spatial resolution in terms of its difference with the next highest spatial
resolution. We find that between time $t - t_{22}^{\rm peak}=-4000 M$ and merger
time $t-t_{22}^{\rm peak}=0$, the amplitude and phase differences decrease with
increasing resolution, as expected, with the medium and high resolution
differing in amplitude by \response{$0.01\%$} at time of merger 
\response{and about $0.1\%$ throughout the inspiral}. During the window between
$t - t_{22}^{\rm peak}=-4000 M$ and $t = t_{22}^{\rm peak}$, the medium and high
spatial resolutions accumulate \response{$0.03$} radians of phase error. After
merger time, the fractional amplitude and phase errors grow, which is expected,
because the amplitude itself is exponentially falling to zero, making
determining the phase accurately increasingly challenging.

\section{Conclusion\label{sec:conclusion}}

We present the first inspiral-merger-ringdown simulations of two binary black
holes using \response{DG} methods. We use the open-source numerical
relativity code \texttt{SpECTRE}~\cite{spectrecode_2024} to perform all
simulations. These include several long-term stability and accuracy tests, e.g.
evolutions of a single black hole in Kerr-Schild, harmonic, and damped
harmonic gauge, as well as a long-term stable gauge wave simulation. All
simulations demonstrate the expected exponential convergence.

The binary black hole simulation is of the last 18 orbits before merger of an
equal mass non-spinning binary. We extract gravitational waveforms at future
null infinity using \texttt{SpECTRE}'s CCE module. We observe exponential
convergence in the constraint violations with increasing resolution, and
demonstrate convergence in amplitude and phase of the $\ell=m=2$ mode of the
gravitational wave strain. \response{The simulations presented here are the
first binary merger simulations where the initial data, evolution, and wave
extraction are all performed using the open-source code \texttt{SpECTRE}.}

\response{Our medium and high resolution \texttt{SpECTRE} simulations run at 80
and 41 $M$/hour on ten 56-core Intel Cascade Lake nodes during inspiral. A
proper comparison between this performance from \texttt{SpECTRE} and an
analogous calculation in \texttt{SpEC} would be nontrivial, in part because a
meaningful comparison should ensure that the \texttt{SpEC} and \texttt{SpECTRE}
simulations have the same accuracy. We plan to perform such a comparison in the
future. Here, for an initial ballpark comparison, we simply note that a
\texttt{SpEC} evolution of analogous initial data, at what we typically would
consider high resolution in \texttt{SpEC} (approximately 300,000 grid points,
with \texttt{SpEC}'s adaptive mesh refinement algorithm varying the precise
number of points throughout the simulation), ran at
approximately $90 M$/hour on 32 Intel Sky Lake cores. The \texttt{SpECTRE}
high-resolution calculation ran about half as fast in wall time, using about 17
times more cores on a grid with about 17 times more gridpoints.
\texttt{SpECTRE}'s domain has many more elements with fewer points per element,
enabling it to scale to more CPU cores than \texttt{SpEC}; however,
\texttt{SpEC}'s choice to use fewer elements with higher number of points
(especially spherical shells near the horizons and in the wave zone) mean that
\texttt{SpEC}'s domain is currently much more efficient. 

We expect that \texttt{SpECTRE} would outperform \texttt{SpEC} in wall time at
sufficiently high resolutions, running on enough CPU cores, but that it would be
less efficient, requiring more cost for the same accuracy. We also expect that
planned future optimizations (discussed in the next paragraph) will greatly
reduce the computational cost of a \texttt{SpECTRE} binary-black-hole
calculation at a given accuracy. Finally, note that in its current state,
\texttt{SpECTRE} is still efficient enough to perform an $\sim 18$ orbit
inspiral---longer than almost all inspirals published to date using
moving-puncture codes---at feasible cost (less than 9 days of wall time on less
than 600 compute cores).}

While the results here present a milestone for \texttt{SpECTRE} simulations,
several further advancements are necessary to enable building catalogs for
future gravitational wave detectors. These fall in one of four categories: i)
performance improvements like using adaptive mesh refinement, dynamic load
balancing, and GPU support; ii) robustness and parameter space improvements like
ensuring that high-spin, high-mass-ratio, and eccentric simulations can be
performed robustly without hand tuning; iii) automation infrastructure that
allows a single user to run hundreds of simulations, such as automatically
restarting failed simulations, automatically transitioning from inspiral to
ringdown, and automatically running CCE after the Cauchy simulation completes;
and iv) improving documentation and tutorials to make the code more accessible
to the broader community. \response{Item i) also includes developing a good
understanding of how error diagnostics, like constraint violations, impact
gravitational-wave amplitude and phase errors; our experience with \texttt{SpEC}
suggests that the answer is complicated and will require careful investigation.}
These will allow \texttt{SpECTRE} to outperform our current code \texttt{SpEC}
and to be more useful to the broader numerical-relativity community.

\ack

Charm++/Converse~\cite{laxmikant_kale_2020_3972617} was developed by the
Parallel Programming Laboratory in the Department of Computer Science at the
University of Illinois at Urbana-Champaign. The figures in this article were
produced with \texttt{matplotlib}~\cite{Hunter:2007,
  thomas_a_caswell_2020_3948793}, \texttt{TikZ}~\cite{tikz}, 
\texttt{ParaView}~\response{\cite{paraview2, paraview}},
\texttt{numpy}~\cite{harris2020array}, \texttt{scipy}~\cite{2020SciPy-NMeth},
and \texttt{scri}~\cite{Boyle:2013nka, Boyle:2014ioa, Boyle:2015nqa,
  mike_boyle_2020_4041972}. Computations were performed at the Resnick
High-Performance Computing Center; a facility supported by the Resnick
Sustainability Institute; at Caltech, the mbot cluster at
Cornell, the ocean cluster at Cal State Fullerton, and
the Urania cluster at the Max Planck Computing and Data Facility.
We are pleased to thank Josh Smith for helpful discussions.
This work was supported
in part by NSF awards PHY-2208014 and AST-2219109, the Dan Black Family Trust,
and Nicholas and Lee Begovich at Cal State Fullerton. This material is based
upon work supported by the National Science Foundation under Grants
No. PHY-2407742, No. PHY- 2207342, and No. OAC-2209655 at Cornell. Any opinions,
findings, and conclusions or recommendations expressed in this material are
those of the author(s) and do not necessarily reflect the views of the National
Science Foundation. This work was supported by the Sherman Fairchild Foundation
at Cornell. Support for this work was provided by NASA through the NASA Hubble
Fellowship grant number HST-HF2-51562.001-A awarded by the Space Telescope
Science Institute, which is operated by the Association of Universities for
Research in Astronomy, Incorporated, under NASA contract NAS5-26555.  This work
was supported in part by the Sherman Fairchild Foundation and by NSF Grants
No.~PHY-2309211, No.~PHY-2309231, and No.~OAC-2209656 at Caltech.
P.K.~acknowledges support of the Department of Atomic Energy, Government of
India, under project no.~RTI4001, and by the Ashok and Gita Vaish Early Career
Faculty Fellowship at the International Centre for Theoretical Sciences.

% IAU suggested abbreviations
\newcommand\aap{Astron.~Astrophys.~}
\newcommand\mnras{Mon.~Not.~R.~Astron.~Soc.~}
\newcommand\prd{Phys.~Rev.~D }

\section*{References}
\bibliographystyle{unsrt}
\bibliography{refs}
\end{document}